\documentclass[12pt]{article}

\usepackage{latexsym,amsmath,amscd,amssymb,graphics}
\usepackage{enumerate,framed,color}
\usepackage{sidecap} 
\usepackage{graphicx}

\usepackage[colorlinks]{hyperref}
\usepackage{url}

\usepackage[all]{xy}

\usepackage[small,bf]{caption}

\usepackage[margin=2cm]{geometry}

\newcommand{\bfib}{\color{blue}\bfseries\itshape}

\newcommand{\rem}[1]{}
\makeatletter

\@addtoreset{figure}{section}
\def\thefigure{\thesection.\@arabic\c@figure}
\def\fps@figure{h, t}
\@addtoreset{table}{bsection}
\def\thetable{\thesection.\@arabic\c@table}
\def\fps@table{h, t}
\@addtoreset{equation}{section}

\makeatother

\rem{%%%%%%%%%%%%%%%%%%%%
\textwidth 6.5 truein
\oddsidemargin 0 truein
\evensidemargin .2 truein
\topmargin -.6 truein
\textheight 9.1 in
}%%%%%%%%%%%%%%%%%%%%

\newcommand{\todo}[1]{\vspace{5 mm}\par \noindent
\framebox{\begin{minipage}[c]{0.95 \textwidth}
\tt #1 \end{minipage}}\vspace{5 mm}\par}
\begin{document}

\newtheorem{theorem}{Theorem}[section]
\newtheorem{definition}[theorem]{Definition}
\newtheorem{lemma}[theorem]{Lemma}
\newtheorem{remark}[theorem]{Remark}
\newtheorem{proposition}[theorem]{Proposition}
\newtheorem{corollary}[theorem]{Corollary}
\newtheorem{example}[theorem]{Example}

\def\below#1#2{\mathrel{\mathop{#1}\limits_{#2}}}

\def\vn{{\vec\nabla}}
\def\he{h_E}
\def\ho{h_O}
\def\ve{v_E}
\def\vo{v_O}
\def\ce{c_E}
\def\co{c_O}
\def\ae{a_E}
\def\ao{a_O}
\def\sio{\psi_O}
\def\sie{\psi_E}
\def\fie{\varphi_E}
\def\fio{\varphi_O}
%********
\def\be{\begin{equation}}
\def\ee{\end{equation}}
\def\bea{\begin{eqnarray}}
\def\eea{\end{eqnarray}}
\def\ba{\begin{array}}
\def\ea{\end{array}}
\def\cA{{\mathcal A}}
\def\cB{{\cal B}}
\def\cC{{\mathcal C}}
\def\cD{{\mathcal D}}
\def\cE{{\mathcal E}}
\def\cF{{\mathcal F}}
\def\cG{{\mathcal G}}
\def\cH{{\mathcal H}}
\def\cI{{\mathcal I}}
\def\cJ{{\mathcal J}}
\def\cK{{\mathcal K}}
\def\cM{{\mathcal M}}
\def\cN{{\mathcal N}}
\def\cO{{\mathcal O}}
\def\cP{{\mathcal P}}
\def\cQ{{\mathcal Q}}
\def\cR{{\mathcal R}}
\def\cS{{\mathcal S}}
\def\cT{{\mathcal T}}
 \def\cU{{\mathcal U}}
\def\cV{{\mathcal V}}
\def\cW{{\mathcal W}}
\def\cX{{\mathcal X}}
\def\cY{{\mathcal Y}}
\def\cZ{{\mathcal Z}}
\def\bOm{\boldsymbol{\Omega}}
\def\hbOm{\widehat{\boldsymbol{\Omega}}}
\def\boldeta{\boldsymbol{\eta}}
\def\brho{\boldsymbol{\rho}}
\def\bM{{\bf M}}
\def\hbM{\widehat{{\bf M}}}
\def\hM{\widehat{M}}

\def\iw{\mbox{\boldmath $i$}\omega}
\def\bi{\mbox{\boldmath $i$}}
\def\s{\sigma}
\def\l{\lambda}
\def\o{\omega}
\def\r{\rho}
\def\L{\Lambda}
\def\D{\Delta}
\def\g{\gamma}
\def\t{\theta}
\def\m{\mu}
\def\a{\alpha}
\def\b{\beta}
\def\e{\epsilon}
\def\ep{\varepsilon}
\def\d{\delta}
\def\n{\nu}
\def\p{\phi}
\def\pw{\partial w}
\def\pt{\partial t}
\def\px{\partial x}
\def\pO{\partial\Omega}
\def\G{\Gamma}
\def\O{\Omega}
\def\xtn{\widetilde{x}_n}
\def\dbR{{\mathop{\rm l\negthinspace R}}}

\def\bbC{\Bbb C}
\def\bbN{\Bbb N}
\def\bbR{{\bf R}}
\def\bbT{\Bbb T}
\def\bbZ{\Bbb Z}
\def\bbS{\Bbb S}
\def\bbG{{\Bbb G}}
\def\bbI{\Bbb I}
\def\BOU{{\bold U}}
\def\BOF{{\bold F}}
\def\BOG{{\bf G}}
\def\BOH{{\bf H}}
\def\bob{{\bf b}}
\def\boc{{\bf c}}
\def\bog{{\bf g}}
\def\bof{{\bold f}}
\def\bou{{\bf u}}
\def\boh{{\bf h}}
\def\borho{{\bf p}}
\def\BOD{{\bf D}}
\def\BOL{{\bf L}}
\def\BOB{{\bf B}}
\def\BOK{{\bf G}}
\def\bow{{\bf w}}
\def\bov{{\bf v}}
\def\boz{{\bf z}}
\def\bga{{\mbox{\boldmath$\gamma$}}}
\def\boy{{\bold y}}
\def\dive{\operatorname{div\/}}
\def\diit{\text{ {\it div\/} }}
\def\diag{\mbox{ diag\,}}
\def\lint{{\int\limits}}
\def\raw{\rightarrow}
\def\lraw{\leftrightarrow}
\def\pa{\partial}
\def\rarp{{x}}
\def\bx{{\mathbf {x} }}
\def\div{\mbox{div}\,}
\def\epi{\epsilon}
\def\vare{\varepsilon}
\def\mum{\nu}
\def\Dal{{\text{\!\!\!\!\!\!\qed}}}
\def\eput#1{\put{ #1}}
\let\<\langle
\let \>\rangle
\def\skp{\vskip 9pt}
\def \qq {\pgothfamily q}
\def \bgamma {\mathbf{\gamma}}
\def \bomega {\mathbf{\omega}}

\newcommand{\remfigure}[1]{#1}%TURNS ON FIGURES
\newcommand{\tphi}{\tilde{\phi}}
\newcommand{\tdelta}{\tilde{\delta}}
\newcommand{\tI}{\tilde{I}}
\newcommand{\tR}{\tilde{R}}
\newcommand{\tv}{\tilde{v}}
\newcommand{\talpha}{\tilde{\alpha}}
\newcommand{\txi}{\tilde{\xi}}
\newcommand{\dR}{\dot{R}}
\newcommand{\dv}{\dot{v}}
\newcommand{\dmu}{\dot{\mu}}
\newcommand{\dbeta}{\dot{\beta}}
\newcommand{\dtR}{\dot{\tilde{R}}}
\newcommand{\dtv}{\dot{\tilde{v}}}
\newcommand{\de}{\delta}
\newcommand{\rhobar}{\overline{\rho}}
\newcommand{\kappabar}{\overline{\kappa}}
\newcommand{\bA}{\boldsymbol{A}}
\newcommand{\bS}{\boldsymbol{S}}
\newcommand{\bm}{\boldsymbol{m}}
\newcommand{\bq}{\boldsymbol{q}}
\newcommand{\bu}{\boldsymbol{u}}
\newcommand{\bPsi}{\boldsymbol{\Psi}}
\newcommand{\bv}{\boldsymbol{v}}
\newcommand{\bX}{\boldsymbol{X}}
\newcommand{\bK}{\boldsymbol{K}}
\newcommand{\bw}{\boldsymbol{w}}
\newcommand{\bphi}{\boldsymbol{\phi}}
\newcommand{\br}{\boldsymbol{r}}
\newcommand{\bGam}{\boldsymbol{\Gamma}}
\newcommand{\bom}{\boldsymbol{\omega}}
\newcommand{\bgam}{\boldsymbol{\gamma}}
\newcommand{\bsigma}{\boldsymbol{\Sigma}}
\newcommand{\bpsi}{\boldsymbol{\Psi}}
\newcommand{\balpha}{\boldsymbol{\alpha}}
\newcommand{\bbeta}{\boldsymbol{\beta}}
\newcommand{\bxi}{\boldsymbol{\xi}}
\newcommand{\bb}{\boldsymbol{b}}
\newcommand{\bF}{\boldsymbol{F}}
\newcommand{\bG}{\boldsymbol{G}}
\newcommand{\bU}{\boldsymbol{U}}
\newcommand{\bd}{\boldsymbol{d}}
\newcommand{\bV}{\boldsymbol{V}}
\newcommand{\bZ}{\boldsymbol{Z}}
\newcommand{\bY}{\boldsymbol{Y}}
\newcommand{\bW}{\boldsymbol{W}}
\newcommand{\bmu}{\boldsymbol{\mu}}
\newcommand{\bPi}{\boldsymbol{\Pi}}
\newcommand{\bXi}{\boldsymbol{\Xi}}
\newcommand{\bTheta}{\boldsymbol{\Theta}}
\newcommand{\bPhi}{\boldsymbol{\Phi}}
\newcommand{\bT}{\boldsymbol{T}}
\newcommand{\bN}{\boldsymbol{N}}
\newcommand{\bkappa}{\boldsymbol{\kappa}}
\newcommand{\bpi}{\boldsymbol{\pi}}
\newcommand{\bSigma}{\boldsymbol{\Sigma}}

%Differential Operators
\newcommand{\pp}[2]{\frac{\partial #1}{\partial #2}}
\newcommand{\dd}[2]{\frac{d #1}{d #2}}
\newcommand{\dede}[2]{\frac{\delta #1}{\delta #2}}
\newcommand{\prt}{\partial}
\newcommand{\DD}[2]{\frac{D #1}{D #2}}

%Greek Symbols
\newcommand{\om}{\omega}
\newcommand{\al}{\alpha}
\newcommand{\da}{\dagger}
\newcommand{\ka}{\kappa}
\newcommand{\ga}{\gamma}
\newcommand{\Om}{\Omega}
\newcommand{\sig}{\sigma}

%Brackets
\newcommand{\lsb}{\left[}
\newcommand{\rsb}{\right]}
\newcommand{\lbb}{\left \langle \left\langle}
\newcommand{\rbb}{\right \rangle\right \rangle}
\newcommand{\lp}{\left(}
\newcommand{\rp}{\right)}
\newcommand{\lb}{\left \langle}
\newcommand{\rb}{\right \rangle}
\newcommand{\lform}[2]{{\big( {#1} \big|\, {#2}\big)}}
\newcommand{\Lform}[2]{{\Big( {#1} \Big|\, {#2}\Big)}}
\newcommand{\scp}[2]{{\left\langle {#1}\, , \, {#2}\right\rangle}}

%Caligraphic Letters
\newcommand{\CX}{{\mathcal X}}
\newcommand{\CO}{{\mathcal O}}
\newcommand{\CL}{{\mathcal L}}
\newcommand{\CH}{{\mathcal H}}
\newcommand{\CA}{{\mathcal A}}
\newcommand{\CF}{{\mathcal F}}
\newcommand{\cL}{{\cal L}}

%Useful operators
\newcommand{\bt}{{\blacktriangle}}
\newcommand{\di}{{\diamond}}

%Set Symbols
\newcommand{\mR}{{\mathbb{R}}}
\newcommand{\mC}{{\mathbb{C}}}
\newcommand{\mH}{{\mathbb{H}}}
\newcommand{\mCP}{{\mathbb{CP}}}
\newcommand{\mg}{{\mathfrak g}}

%Lie Groups/Algebras
\newcommand{\Ad}{\mbox{Ad}}
\newcommand{\ad}{\mbox{ad}}
\newcommand{\msu}{\mathfrak{su}}
\newcommand{\mse}{\mathfrak{se}}
\newcommand{\mso}{\mathfrak{so}}

%Misc
\newcommand{\id}{{\mathrm{id}}\,}
\newcommand{\ti}{\times}
\newcommand{\tr}{\mbox{tr}}
\newcommand{\im}{\mbox{im}}
\newcommand{\non}{\nonumber\\}
\newcommand{\con}{\overline}
\newcommand{\cst}{\text{cst}}
\newcommand{\sech}{\text{sech}}
\newcommand{\bra}[1]{\left \langle #1 \right |}
\newcommand{\ket}[1]{\left | #1 \right \rangle}
\newcommand{\hor}{\mbox{Hor}}
\newcommand{\ver}{\mbox{Ver}}

%Half angle trig.
\newcommand{\hc}[1]{{\cos \frac{#1}{2}}}
\newcommand{\hs}[1]{{\sin \frac{#1}{2}}}

%Headings
\newtheorem{thm}{Theorem}[section]
\newtheorem{cor}[thm]{Corollary}
\newtheorem{defi}[thm]{Definition}
\newtheorem{prop}[thm]{Proposition}
\newtheorem{lem}[thm]{Lemma}
\newtheorem{rema}[thm]{Remark}

\newcommand{\comment}[1]{\vspace{1 mm}\par
\marginpar{\large\underline{}}\noindent
\framebox{\begin{minipage}[c]{0.95 \textwidth}
{\bfib #1} \end{minipage}}\vspace{1 mm}\par}
%

%%%%%%%%%%%%%%%%%%%%%%%%%%%%%%%%%%%%%%%%%%%%%%%%%%%%%%%%%%%%%%%%%%%%%%%%%%%%%%%
%%%%%%%%

%%%%%%%%%%%%%%%%%%%%%%%%%%%%%%%%%%%%%%%%%%%%%%%%%%%%%%%%%%%%%%%%%%%%%%%%%%%%%%%
%%%%%%%%

\title{Exact geometric theory of dendronized polymer dynamics}
\author{Fran\c{c}ois Gay-Balmaz$^{1}$, Darryl D. Holm$^{2}$,\\  Vakhtang Putkaradze$^{3,4}$, Tudor S. Ratiu$^{5}$}
\addtocounter{footnote}{1}
\footnotetext{Control and Dynamical Systems, California Institute of Technology 107-81, Pasadena, CA 91125, USA and Laboratoire de 
M\'et\'eorologie Dynamique, \'Ecole Normale Sup\'erieure/CNRS, Paris, France. 
\texttt{fgbalmaz@cds.caltech.edu}
\addtocounter{footnote}{1} }
\footnotetext{Department of Mathematics and Institute for Mathematical Sciences, Imperial College, London SW7 2AZ, UK. 
\texttt{d.holm@imperial.ac.uk}
\addtocounter{footnote}{1} }
\footnotetext{Department of Mathematics,
Colorado State University,
Fort Collins, CO 80523-1874, USA.
\texttt{putkarad@math.colostate.edu}
\addtocounter{footnote}{1}}
\footnotetext{Department of Mechanical Engineering,
University of New Mexico,  Albuquerque, NM 87131-1141, USA.
\addtocounter{footnote}{1}}
\footnotetext{Section de
Math\'ematiques and Bernoulli Center, \'Ecole Polytechnique F\'ed\'erale de
Lausanne,
CH--1015 Lausanne, Switzerland.
\texttt{tudor.ratiu@epfl.ch}
\addtocounter{footnote}{1} }
\date{May 13, 2010}

\maketitle

\makeatother

\maketitle

%|||-------------------text width----------------------|||

%-------------------------------------------------------
%--------------------------------------------------------

\abstract{  
Dendronized polymers consist of an elastic backbone with a set of iterated branch structures (dendrimers) attached at every base point of the backbone. The conformations of such molecules depend on the elastic deformation of 
the backbone and the branches, as well as on nonlocal (\emph{e.g.}, electrostatic, or Lennard-Jones) interactions between the elementary molecular units comprising the dendrimers and/or backbone.  
We develop a geometrically exact theory for the dynamics of such
polymers, taking into account both local (elastic) and nonlocal interactions. 
The theory is based on applying symmetry reduction of Hamilton's principle for a Lagrangian defined on the tangent 
bundle of iterated semidirect products of the rotation groups that represent the
relative orientations of the dendritic branches of the polymer. 
The resulting symmetry-reduced equations of motion are written in conservative form. 
} 

\tableofcontents
\section{Physical and mathematical motivations} 
Dendronized (dendritic) polymers represent an exciting  recent  advance  in the field of chemical synthesis. These compound molecular structures are formed by assembling multiple \emph{dendrimers} (a low molecular weight unit to which a number of dendrons, or branches, is attached) that are each connected by its base to a long polymeric backbone. Known as \emph{rod-shaped dendrimers}, the first patent for synthesizing them was filed in 1987 \cite{Tomalia-etal-1987}. The description of dendronized polymers appeared for the first time in the scientific literature in 1990 \cite{ToNaGo1990}. Since then, the field of dendronized polymers has grown explosively, as a great number of papers have been published on the subject. For discussions of recent progress, literature reviews, and methodology we refer to \cite{Schluter1998,Fr2005,Ro-etal-2009}. 

Although the literature describing experimental approaches (such as fabrication and measurement) of dendronized polymers is extensive, much less work has been devoted to modeling {their dynamical} properties. In our opinion, this dearth of modeling is probably due to the mathematical difficulties posed by the analysis of simultaneous forces caused by the elastic deformations of the polymer backbone and its attached dendrimers, as well as long-range interactions between the dendrimers through screened electrostatic and other forces. From the mathematical point of view, obstacles arise in the extension of the classical and well-developed Kirchhoff theory of elastic rods \cite{DiLiMa1996}, mainly because the description of dendronized polymers is geometrically considerably more complex and the long-range interactions among different parts of the polymers. We shall note, however, that there has been some recent progress in modeling dendronized polymer conformations. For example, the course-grained approach \cite{Chr-etal-2006} and the atomistic/Janus chain model \cite{Di-etal-2007} have been used to study the formation of helical structures. In addition, Monte-Carlo simulations were used to predict 
stiffness of polymer chains in \cite{Co-etal-2005}.   Nevertheless, we are aware of no previous work that is capable of describing the \emph{spatio-temporal } evolution of the dendronized polymer. This evolutionary aspect is important, for example, in the computation of the propagation of sound waves along the polymer, or in choosing a dynamical route toward a final configuration.

\begin{figure}
%\begin{SCfigure}
\centering
%\captionsetup{width=0.8 \textwidth}
\includegraphics[width=0.4\textwidth,angle=0]{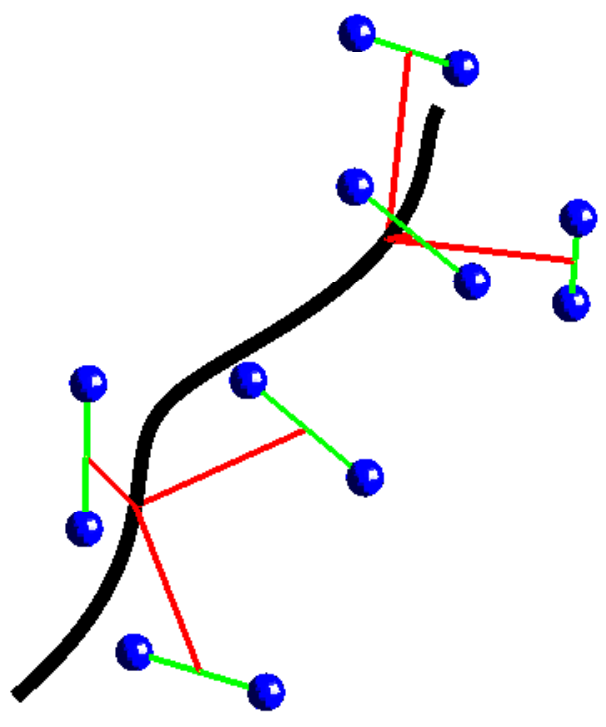}
\setlength{\captionmargin}{45pt}
\caption{\small{Motivating physical system for the paper: a sketch of dendronized polymer. Rigid conformations of charges (red) are distributed along the centerline (solid black curve).  At the end of each branch in a given rigid bouquet, \emph{another} rigid bouquet of charges is attached (denoted here by two blue spheres on a rigid green rod). The orientations of the bouquets attached at the ends of different branches of the same dendrite (red bouquet) may, in general, be different. More bouquets can then be attached to the blue spheres, and so forth. The system experiences elastic forces induced by deformations of the centerline and the relative deformation of bouquets, as well as electrostatic and other nonlocal forces (e.g., Lennard-Jones) due to the interactions between each pair of charges.  
}}
%\end{SCfigure}
 \label{fig:multi-bouquet}
\end{figure}

This paper aims to derive the general and geometrically exact theory for dynamical evolution of dendronized molecules from fundamental physical principles. Our theory can deal with  \emph{arbitrary elastic and nonlocal  interactions} arising from charge distributions on branched structures and is thus applicable to a wide range of dendronized polymers. The system we have in mind is illustrated on Figure~\ref{fig:multi-bouquet}: rigid \emph{bouquets} of charges are attached to each point of the deformable elastic centerline. In addition, another set of rigid bouquets is attached to each end of the first bouquet. We derive equations of motion in a general geometric setting for an arbitrary number of sequential bouquets. However, when transferring these equations from their general geometric form to the ``nuts and bolts" vector equations for angular and linear momenta of each bouquet, these equations become algebraically quite complicated if the number of charge bouquets in the sequence is greater than two.

The description of such a compound system will be made possible by using iterated semidirect products of Lie groups. The equations of motion are then derived by using a Lagrangian reduction {by symmetry} inspired by the standard Euler-Poincar\'e reduction on Lie groups \cite{HoMaRa1998}. 

\paragraph{Euler-Poincar\'e reduction.}
Euler-Poincar\'e reduction is a useful tool for deriving reduced Euler-Lagrange equations when a variational principle admits a Lie group of symmetries. In the simplest case, one starts with a Lagrangian $L:TG\rightarrow\mathbb{R}$ defined on the tangent bundle $TG$ of a Lie group $G$ and assumes that it is invariant under translation by $G$ acting from the left, for example. This Lie symmetry allows one to define a reduced Lagrangian $\ell:\mathfrak{g}\rightarrow\mathbb{R}$ on the Lie algebra $\mathfrak{g}=TG/G$ of $G$. Hamilton's variational principle $\delta \int_{t_0}^{t_1}L(g,\dot g)dt=0$ for the left $G$-invariant Lagrangian $L(g,\dot g)=L(hg,h\dot g)$ for any $h\in G$ implies a variational principle $\delta\int_{t_0}^{t_1}\ell(\xi)dt=0$ with $\ell(\xi):=L(e,g^{-1}\dot g)$ for the reduced Lagrangian, provided the  variations of $\xi:=g^{-1}\dot g\in\mathfrak{g}$ are of the form $\delta\xi=\dot\eta+[\xi,\eta]$ for some
curve $\eta \in \mathfrak{g}$ satisfying $\eta(t_0) = \eta(t_1) = 0 $. From this equivalence, one shows that the Euler-Lagrange equations are equivalent to the Euler-Poincar\'e equations
\[
\frac{d}{dt}\frac{\delta \ell}{\delta\xi}=\operatorname{ad}^*_\xi\frac{\delta\ell}{\delta\xi},
\]
where $\xi=g^{-1}\dot g$. 
This formulation is useful for the evolutionary description of rigid bodies and incompressible flows, for example. In \cite{HoMaRa1998} and \cite{GBRa2009}, this approach has been generalized to include the case of systems defined on semidirect products of Lie groups with cocycles in order to
gain geometric insight on \emph{advected quantities}. Systems treated by this theory include heavy tops, compressible fluids, complex fluids, and liquid crystals, for example. As we will show, the description of the dynamics of multi-bouquets summons a more sophisticated version of Euler-Poincar\'e reduction involving iterated semidirect products.

\section{Review of the geometry of molecular strands}

In this section, we briefly recall the theory of affine Euler-Poincar\'e reduction and its application to the dynamics of molecular strands that do not have a dendritic structure.  Our exposition follows the original \emph{exact geometric rod theory} \cite{SiMaKr1988} that has been 
adapted to molecular strands with nonlocal interactions in \cite{HoPu2009,ElGBHoPuRa2010}. 

\subsection{Affine Euler-Poincar\'e reduction}
\label{subsec_strand_review}

Let $G $ be a Lie group. For the molecular strand $G=SE(3)$, the special Euclidean group of orientation preserving rotations and translations in three-dimensional space. Let $V $ be a left representation 
space for $G $ and let $c:G \rightarrow V ^\ast$ be a \emph{group one-cocycle}. That is, $c$ satisfies
\[
c(gh) = g c(h) + c(g)
\,,
\]
for all $g, h \in G$. Suppose $a \in V^\ast$ and denote by 
\[
\theta_g(a): = ga + c(g)
\]
the \emph{induced affine action} of $G $ on $V ^\ast$. Recall
that $ga \in V ^\ast$ is defined by $\left\langle ga , v \right\rangle = \left\langle a , g ^{-1} v \right\rangle$, 
for all $v \in V $. 

 Consider a Lagrangian $L_{a_0}:TG\rightarrow\mathbb{R}$ depending on the parameter $a_0\in V^*$ such that the function 
 \[
 L:TG\times V^*\rightarrow\mathbb{R}
 ,\quad
 L(g,\dot g, a_0):=L_{a_0}(g,\dot g)
 \]
 is $G$-invariant. Then $\ell: \mathfrak{g}\times V^* \rightarrow \mathbb{R}$ given by
 \[
\ell(\xi,\theta_g(a_{0})):=L(g^{-1}\xi,a_0)
\]
is well defined. Hamilton's variational principle 
\[ \delta\int_{t_0}^{t_1}L_{a_0}(g,\dot g)dt=0 
\] 
 reduces then to the constrained variational principle 
 \[ 
 \delta\int_{t_0}^{t_1}\ell(\xi,a)dt=0,
 \]  
for special variations $\delta\xi \in \mathfrak{g}$ and $\delta a \in V ^\ast$ that are obtained as follows. If 
we denote  $\xi=g^{-1} \dot g\in \mathfrak{g}$, then a direct computation shows that
 \[ 
 \de \xi= \de (g^{-1} \dot g) 
 = - g^{-1} \de g\, g^{-1} \dot g + g^{-1} \de \dot g
 =: \frac{d \eta}{d t}+\big[   \xi \, , \, \eta \big] 
 = \frac{d \eta}{d t}+ \operatorname{ad}_\xi  \eta
\,,
 \] 
where $\eta = g^{-1} \de g \in \mathfrak{g}$ and $[\, \cdot \,,\, \cdot \,]$ is the Lie bracket in $\mathfrak{g}$, the Lie algebra of $G$. The variation in $a\in V^*$ may be obtained from the path $a(\epsilon) := \theta_{g^{-1} (\epsilon)} a_0 \in V ^\ast$, where $g( \epsilon) \in G $
is a path with $g(0) = e $ and $g'(0) = \eta \in \mathfrak{g}$. Then  
 \[ 
 \de a =\de  \big( g^{-1} a + c(g^{-1} ) \big) = - \eta a - \mathbf{d} c (\eta) \, , 
 \] 
 where $\mathbf{d}c$ is the derivative of the one-cocycle $c$ at the identity, and $\eta= g ^{-1}\delta g \in \mathfrak{g}$ is an arbitrary curve on $[t_0,t_1]$ such that $\eta(t_0) = 
 \eta(t_1) = 0 $.  This symmetry-reduced variational principle leads to the \emph{affine Euler-Poincar\'e equations}
\begin{equation}\label{general_affine_equ}
\left\{
\begin{array}{l}
\vspace{0.2cm}\displaystyle\frac{\partial}{\partial t}\frac{\delta
\ell}{\delta\xi}=\operatorname{ad}^*_\xi\frac{\delta \ell}{\delta\xi}+\frac{\delta
l}{\delta a}\diamond a-\mathbf{d}c^T\left(\frac{\delta \ell}{\delta a}\right),\\
\displaystyle\frac{\partial}{\partial t} a + \xi a + \mathbf{d}c( \xi) = 0,
\end{array}
\right.
\end{equation}
see \cite{GBRa2009}. Here $\diamond: V \times V ^\ast 
\rightarrow \mathfrak{g}^\ast$ is defined by 
$\left\langle u \diamond b, \zeta \right\rangle: = 
-\left\langle \zeta b, u \right\rangle : =\left\langle b, \zeta u \right\rangle$ for all $\zeta\in \mathfrak{g}$, $u \in V $, $b \in V ^\ast$, where $\zeta b \in \mathfrak{g}$ denotes
the infinitesimal $\mathfrak{g}$-representation on $V$ induced by the given $G$-representation.

 The last two terms in the first equation of the system \eqref{general_affine_equ} come from variations in $a$ as follows: 
\[ 
\left\langle\frac{\de l}{\de a} , \de a 
\right\rangle
= 
\left\langle
\frac{\de l}{\de a} ,  - \eta a - \mathbf{d} c (\eta) 
\right\rangle 
= 
\left\langle
\frac{\de l}{\de a} \diamond a- \mathbf{d} c^T \left(\frac{\de l}{\de a}  \right) , \eta  \right\rangle . 
\] 

For the molecular strand, a  generalization is needed, namely, we suppose that the Lagrangian $L_{a_0}$ is known only for a \textit{fixed} $a_0\in V^*$. In particular, we do not
suppose that this Lagrangian comes from a $G$-invariant 
function  $L:TG\times V^*\rightarrow\mathbb{R}$. In this case, we assume that $L_{a_0}: TG \rightarrow \mathbb{R}$ is only invariant under the isotropy subgroup 
$G^c_{a_0}: = \{ g \in G \mid \theta_g (a_0) = a_0\}$ of $a_0$, that is, 
\[
L_{a_0}(hg, h\dot{g}) = L_{a_0}(g, \dot{g})
\]
for all $h \in G^c_{a_0}$.  The reduction goes as before, $\ell(\xi,\theta_g(a_{0}))=L_{a_{0}}(g^{-1}\xi)$, except that now the reduced Lagrangian $\ell$ is only defined on the submanifold $\mathfrak{g}\times\mathcal{O}^c_{a_{0}}\subset \mathfrak{g}\times V ^\ast$,
where $\mathcal{O}^c_{a_{0}}:=
\{\theta_g(a_{0})\mid g\in G\} \subset V ^\ast$ is the 
$G $-orbit of $a_0$. The associated quotient map is, as before,
\begin{equation}\label{quotient_AEP}
(g,\dot g)\in TG \mapsto (g^{-1}\dot g,c(g^{-1})) \in \mathfrak{g}\times V ^\ast.
\end{equation}
The reduced equations of motion are still given by the affine Euler-Poincar\'e equations \eqref{general_affine_equ}, where $\ell$ is arbitrarily extended to a smooth function
on $\mathfrak{g} \times  V ^\ast$. It is shown in \cite{ElGBHoPuRa2010} that these equations are independent
of the extension of $\ell$ to $\mathfrak{g} \times V ^\ast$.
Note also that the solution of the second equation in \eqref{general_affine_equ} is $a(t) = 
\theta_{g(t) ^{-1}} a_0 $, where $\xi(t) = g(t) ^{-1} \dot{g}(t)$, $g(0) = e$. 

In many applications $a_0 = 0$. It may also happen that the
reduced Lagrangian $\ell $ cannot be explicitly expressed
as a function of $\xi$ and $a$ only and it has the form
$\ell(\xi, a, g) $, where $g \in G $ is such that $\theta_{ g ^{-1}} a_0 = a $. If we assume that $\ell $ is $G^c_{a_0} $-invariant, that is, $\ell(\xi, a, hg) = \ell(\xi, a, g)$ for all $h \in G^c_{a_0}$, then $\ell( \xi, a, g ) $ is a well defined function of $\xi$ and $a $ and equations 
\eqref{general_affine_equ} are still valid, where one
computes the functional derivatives as if $g $ were expressed
explicitly in terms of $\xi$ and $a $. Even if this is
not possible, the derivatives still have explicit expressions
in many applications, as shown in \cite{ElGBHoPuRa2010} and 
discussed below.

\subsection{Molecular strand dynamics}\label{review_strand}

We briefly review the geometric setting for the molecular
strand from \cite{HoPu2009,ElGBHoPuRa2010}. Denote $I := 
[a, b] \subset \mathbb{R}$ and let $SE(3)$ be the special Euclidean group of $\mathbb{R}^3$ consisting of orientation preserving rotations and translations. Denote by 
$\mathfrak{se}(3)$ the Lie algebra of $SE(3)$.
The material configuration variables  are the spatial position $\br(s,t)\in\mathbb{R}^3$ of the filament and the rotation $\Lambda(s, t)$ of the rigid charge conformation at parameter value $s$ and time $t$. The configuration space of this system is therefore the  group $\mathcal{F}(I,SE(3))$ of $SE(3)$-valued smooth mappings on $I$. The multiplication in $\mathcal{F}(I, SE(3))$ is given by pointwise multiplication in $SE(3)$:
\begin{equation}
\label{se_3_mult}
(\Lambda_1, \br_1)(\Lambda_2, \br_2)
= ( \Lambda_1\Lambda_2, \br_1 + \Lambda_1 \br_2).
\end{equation}

The time and space derivatives yield, respectively, the \textit{material velocity} $(\dot\Lambda(s,t),\dot{\br}(s,t))$ and the angular and linear \textit{deformation gradients} $(\Lambda'(s,t),\br'(s,t))$. Given $\Lambda$ and $\br$, we define the following \textit{reduced convective variables} \cite{SiMaKr1988},
\begin{eqnarray}
\Om&= \,\Lambda^{-1} \Lambda'&  \in \mso(3),\nonumber
\\
\om&=\, \Lambda^{-1} \dot{\Lambda}& \in \mso(3),\nonumber
\\
\bGam&=\,\Lambda^{-1} \br'& \in \mathbb{R}^3,\label{bundle.coords}
\\
\bgam&=\, \Lambda^{-1} \dot{\br}& \in \mathbb{R}^3,\nonumber
\\
\brho&= \,\Lambda^{-1} \br&  \in \mathbb{R}^3.\nonumber
\end{eqnarray}
The physical interpretation of the variables \eqref{bundle.coords} is as follows.  The variable $\brho(s,t)$ represents the \textit{position of the filament in space as viewed by an observer} who rotates with the rigid charge conformation at $(s,t)$. The variables \big($\Om(s,t), \bGam(s,t)$\big) describe the \textit{deformation gradients as viewed by an observer} who rotates with the with the rigid charge conformation.  The variables \big($\om(s,t), \bgam(s,t)$\big) describe the \textit{body angular velocity} and the \textit{linear velocity as viewed by an observer} who rotates with the rigid charge conformation.

Remarkably, the convective variables \eqref{bundle.coords} can be obtained by considering the quotient map
\begin{align}
\label{special_quotient_map}
&T \mathcal{F}(I, SE(3)) \rightarrow \mathcal{F}(I, \mathfrak{se}(3)) \times  \mathcal{F}(I, \mathfrak{so}(3)) \times \mathcal{F}(I, \mathbb{R}^3)^2,\nonumber\\
&(\Lambda,\dot\Lambda,\br,\dot\br)\mapsto (\om,\bga,\Om,\bGam,\brho)=\left((\Lambda,\br)^{-1}(\Lambda,\dot\Lambda,\br,\dot\br),c((\Lambda,\br)^{-1})\right),
\end{align}
where $(\Lambda,\br)^{-1}(\Lambda,\dot\Lambda,\br,\dot\br)$ denotes the tangent lift of left translation on $SE(3)$ and $c$ is the group one-cocycle defined by
\[
c( \Lambda, \br) : = \left((\Lambda, \br) \partial_s(\Lambda, \br)^{-1}, - \br \right).
\]
This is a cocycle with respect to the representation of 
the group $\mathcal{F}(I,SE(3))$ on the vector space 
$\mathcal{F}(I,\mathfrak{se}(3)) \times 
\mathcal{F}(I,\mathbb{R}^3)$ given by
\[
(\Lambda,r)(\Omega,\boldsymbol{\Gamma},\boldsymbol{\rho})=\left(\operatorname{Ad}_{(\Lambda,\boldsymbol{r})}(\Omega,\boldsymbol{\Gamma}),\Lambda\boldsymbol{\rho}\right),
\]
where
$(\Lambda,\br) \in \mathcal{F}(I, SE(3))$, 
$(\Omega, \boldsymbol{\Gamma}) \in \mathcal{F}(I, 
\mathfrak{se}(3)) = \mathcal{F}(I, \mathfrak{so}(3)) 
\times \mathcal{F}(I, \mathbb{R}^3)$, 
$\boldsymbol{\rho}\in \mathcal{F}(I, \mathbb{R}^3)$, and 
$\operatorname{Ad}_{(\Lambda, \br)}$ is the adjoint action
of the element $(\Lambda, \br) \in \mathcal{F}(I, SE(3))$ 
on $\mathfrak{se}(3)$. 
Hence the affine representation of $\mathcal{F}(I, SE(3))$ 
on $\mathcal{F}(I, \mathfrak{se}(3)) \times \mathcal{F}(I, \mathbb{R}^3)$ is given by
\[
\theta_{(\Lambda,\boldsymbol{r})}(\Omega,\boldsymbol{\Gamma},\boldsymbol{\rho})=(\Lambda,\boldsymbol{r})(\Omega,\boldsymbol{\Gamma},\boldsymbol{\rho})+c(\Lambda,\boldsymbol{r}).
\]
It is easy to see that the isotropy group of $({0},\boldsymbol{0},\boldsymbol{0})$ relative to this affine representation is $\mathcal{F}(I, SE(3))^c_{0} = SO(3)$. The projection \eqref{special_quotient_map} is therefore of the general form described in the quotient map \eqref{quotient_AEP}, with $G=\mathcal{F}(I,SE(3))$ and $V^*=\mathcal{F}(I, \mathfrak{se}(3) \times \mathcal{F}(I,\mathbb{R}^3)$.

We consider reduced Lagrangians of the form
\[
\ell(\om,\bga,\Om,\bGam,\brho)=\ell_{loc}(\om,\bga,\Om,\bGam,\brho)+\ell_{np}(\Om,\bGam,\brho,(\Lambda,\br)),
\]
where the first Lagrangian $\ell_{loc}$ is explicitly given in terms of the variables $(\om,\bga,\Om,\bGam,\brho)$ and the second Lagrangian $\ell_{np}$ has still a dependence on $(\Lambda,\br)$, where $(\Lambda,\br)$ are such that $(\Lambda^{-1}\Lambda',\Lambda^{-1}\br',\Lambda^{-1}\br)=(\Om,\bGam,\brho)$. The Lagrangian $\ell_{np}$ is well
defined because it is required to be $SO(3)$-invariant in the following sense. Regard $SO(3) $ as a subgroup of 
$\mathcal{F}(I, SE(3))$, $\Lambda \in SO(3) \mapsto (\Lambda, {\bf 0}) \in  \mathcal{F}(I, SE(3))$. Then  it is assumed that  $\ell_{np}$ is invariant relative to the multiplication \eqref{se_3_mult}, that is,
\begin{equation}
\label{so_3_invariance}
\ell_{np} (\Omega, \boldsymbol{\Gamma}, 
\boldsymbol{\rho}, (h \Lambda, h\br)) = \ell_{np} (\Omega, \boldsymbol{\Gamma}, 
\boldsymbol{\rho}, (\Lambda, \br)),\quad\text{for all}\quad h\in SO(3).
\end{equation}

For the molecular strand, this dependence on $(\Lambda,\br)$ is given in terms of the $SO(3)$-invariant variables $(\xi,\bkappa)$ defined by
\[
(\xi(s,s'),\bkappa(s,s')):=(\Lambda(s),\br(s))^{-1}(\Lambda(s'),\br(s'))
= \left(\Lambda(s)^{-1} \Lambda(s'), \Lambda(s)^{-1} 
(\br(s') - \br(s)) \right)
\]
and we have the concrete expression
\[
\ell_{np}(\Om,\bGam,\brho,(\Lambda,\br))=\iint U(\xi(s,s'),\bkappa(s,s'),\bGam(s),\bGam(s')){\rm d}s{\rm d}s'
\]
for a given function $U: \mathfrak{so}(3) \times \left(\mathbb{R}^3 \right)^3 \rightarrow \mathbb{R}$.
In this expression, the relation \eqref{so_3_invariance} follows by $SO(3)$-invariance of the
variables $\xi(s, s') \in \mathfrak{so}(3)$, $\boldsymbol{\kappa}(s, s')\in \mathbb{R}^3 $.

Let us show that this general form allows us to treat a nonlocal potential energy of interaction depending on the spatial distances between the individual charged units in different locations along the strand. The spatial reference state for the $k$th charge in a given rigid charge conformation is the sum $\br(s)+\boldeta_k(s)$, where $\boldeta_k(s)$ is a vector of constant length. In the current configuration, the position $\boldsymbol{c}_k(s)$ of the charge $k$ is thus $\boldsymbol{c}_k(s)=\br(s)+\Lambda(s)\boldeta_k(s)$ and the distance from charge $k$ at spatial position $\boldsymbol{c}_k(s)$ to charge $m$ at position $\boldsymbol{c}_m(s')$ is 
\[
d_{km}(s,s'):=|\boldsymbol{c}_k(s)-\boldsymbol{c}_m(s')|.
\]
The chain of equalities
\begin{align*}
d_{km}(s,s'):&=|\boldsymbol{c}_k(s)-\boldsymbol{c}_m(s')|=|\br(s)+\Lambda(s)\boldeta_k(s)-\br(s')-\Lambda(s')\boldeta_m(s')|\\
&=|\Lambda(s)^{-1}(\br(s)-\br(s'))+\boldeta_k(s)-\Lambda(s)^{-1}\Lambda(s')\boldeta_m(s')|\\
&=|-\bkappa(s,s')+\boldeta_k(s)-\xi(s,s')\boldeta_m(s')|,
\end{align*}
shows that $d_{km}(s,s')$ can be expressed in terms of the quantities $\xi(s,s')\in \mathfrak{so}(3)$, $\bkappa(s,s')
\in \mathbb{R}^3$, as required.

The equations of motion associated with the reduced Lagrangian $\ell $ are therefore obtained by affine Euler-Poincar\'e reduction for a $SO(3)$-invariant Lagrangian defined on $T\mathcal{F}(I,SE(3))$ and with respect to a reference configuration given by $a_{0}=0$. Using \eqref{general_affine_equ} produces the equations,
\begin{equation}\label{Final_Euler_Poincare_equations}
\left\lbrace\begin{array}{l}
\vspace{.2cm}
\displaystyle\lp \prt_t + \bom\times\rp\dede{\ell}{\bom} 
+ \lp\prt_s + \bOm\times\rp\dede{\ell}{\bOm} 
+\brho\times\dede{\ell}{\brho}+\bGam\times\dede{\ell}{\bGam}
+\bgam\times\dede{\ell}{\bgam}=0,\\ 
\displaystyle\lp \prt_t + \bom\times\rp\dede{\ell}{\bgam} 
+ \lp\prt_s + \bOm\times\rp\dede{\ell}{\bGam}
 -\dede{\ell}{\brho}=0,
\end{array}\right.
\end{equation}
where $\bom$ is the vector in $\mathbb{R}^3$ defined by the 
condition $\om \mathbf{w} =
\bom \times \mathbf{w}$, for any $\mathbf{w}\in\mathbb{R}^3$, and similarly for $\bOm$. In the equations above, the functional derivatives of $\ell$ relative to $(\boldsymbol{\Omega}, \boldsymbol{\Gamma}, \boldsymbol{\rho})$ are taken as if one could have expressed explicitly $\ell$ only in terms of the variables $(\boldsymbol{\omega}, \boldsymbol{\gamma}, \boldsymbol{\Omega}, \boldsymbol{\Gamma}, \boldsymbol{\rho})$. Of course, this dependence
cannot be explicitly written down since there is no
concrete expression of $( \Lambda, \br)$ in terms of  
$(\boldsymbol{\Omega}, \boldsymbol{\Gamma}, \boldsymbol{\rho})$ but, as shown in \cite{ElGBHoPuRa2010}, the derivatives $\delta \ell/ \delta \boldsymbol{\Omega}$, 
$\delta \ell/ \delta \boldsymbol{\Gamma}$, $\delta\ell/ \delta\boldsymbol{\rho}$ have an explicit expression involving only the variables $(\boldsymbol{\omega}, \boldsymbol{\gamma}, \boldsymbol{\Omega}, \boldsymbol{\Gamma}, \boldsymbol{\rho})$. Moreover, the system
\eqref{Final_Euler_Poincare_equations} can be written more explicitly as
\begin{equation}
\left\lbrace\begin{array}{l}
\vspace{.2cm}
\displaystyle
\left(\partial_t 
+ \bom \times \right)\frac{\delta \ell_{loc}}{\delta \bom}+ \left(\partial_s + \bOm \times \right)
\frac{\delta \ell_{loc}}{\delta \bOm}
+ \bgam \times \frac{\delta \ell_{loc}}{\delta \bgam} 
+ \bGam  \times \frac{\delta \lp \ell_{loc}+
\ell_{np} \rp }{\delta \bGam} 
+ \brho \times \frac{\delta \ell_{loc}}{\delta \brho}  
\nonumber \\
\vspace{.2cm}\displaystyle\qquad\qquad\qquad=\int \left( \frac{\partial U}{\partial \bkappa} (s,s') \times \bkappa (s,s') 
+ \mathbf{Z}(s,s') \right) \mbox{d} s',\\
\vspace{.2cm}\displaystyle\left(\partial_t +\bom \times \right)
\frac{\delta \ell_{loc}}{\delta \bgam}
+ \left(\partial_s +\bOm \times \right)
\frac{\delta \lp  \ell_{loc}+\ell_{np} \rp }{\delta \bGam}
- \frac{\delta \ell_{loc}}{\delta \brho}\nonumber\\
\vspace{.2cm}\displaystyle\qquad\qquad\qquad=
\int \left(
\xi(s,s') \frac{\partial U}{\partial \bkappa} (s',s)
-\frac{\partial U}{\partial \bkappa} (s,s')
\right) \mbox{d} s',
\end{array}\right.
\end{equation}
where the term $\bZ(s,s')$ is the vector given by
\begin{equation*}
\widehat\bZ(s,s') = \xi(s,s')
\lp
\frac{\partial U}{\partial \xi} (s,s')
\rp ^T
-\frac{\partial U}{\partial \xi} (s,s')
\xi^T(s,s'),
\end{equation*}
and the functional derivatives of $\ell_{np}$ with respect to the variables $(\boldsymbol{\Omega}, \boldsymbol{\Gamma}, \boldsymbol{\rho})$ are computed in the usual manner. 
For all the details leading to this system see 
\cite{ElGBHoPuRa2010}.

\section{Reduction for iterated semidirect products}
\label{sec_geometric_setting}

In order to extend the theory of molecular strand dynamics to dendronized polymers, we need to utilize additional  mathematical concepts, based on the notion of iterated semidirect products. Physically, the rotation of the $k$th bouquet level induces the motion of all subsequent bouquets starting with the level $k+1$; this can be formulated 
mathematically in terms of iterated semidirect products. 
\subsection{Semidirect product with cocycle}
\label{sec_sdp}

Let $G_1 $ and $G_2 $ be Lie groups and assume that $G_1 $ 
acts on $G_2 $ by Lie group homomorphisms. Define 
the semidirect product
$G_1\,\circledS\,G_2$ with multiplication
\[
(g_1,g_2)(\bar{g}_1,\bar{g}_2):=(g_1\bar g_1,g_2(g_1\cdot \bar g_2)), \qquad g_1, \bar{g}_1 \in G_1, g_2, \bar{g}_2 \in G_2,
\]
where $g_1\cdot \bar g_2 $ denotes the action of $g_1$ on
$\bar{g}_2 $. The identity element is $(e,e)$ and $(g_1, 
g _2) ^{-1}= ( g_1 ^{-1}, g_1 ^{-1}\cdot g_2 ^{-1})$. As 
we shall see later, the choice $G_1 = SE(3)$, $G_2 = SO(3) $, where the semidirect product $SE(3) \,\circledS\,SO(3) $
is defined by the action $(\Lambda_1, \boldsymbol{r}) \cdot \Lambda_2: = \Lambda_1 \Lambda_2 \Lambda_1^{-1}$, $(\Lambda_1,\boldsymbol{r}) \in SE(3)$, $\Lambda_2\in SO(3)$, will be
of great interest in the geometric study of two-level bouquets.

We suppose that $G_1\,\circledS\,G_2$ acts on a dual vector space $V^*$ by an affine action
\[
\theta_{(g_1,g_2)}a=(g_1,g_2)a+c(g_1,g_2), \quad a \in V ^\ast
\]
where $a\mapsto (g_1,g_2)a$ denotes a representation of $G_1\,\circledS\,G_2$ on $V^*$ and $c:G_1\,\circledS\,G_2\rightarrow V^*$ is a group one-cocycle with respect to the above representation, that is,
\[
c((g_1,g_2)(\bar{g}_1,\bar{g}_2))=(g_1,g_2)c(\bar{g}_1,\bar{g}_2) + c(g_1,g_2)
\]
for all $(g_1,g_2), (\bar{g}_1,\bar{g}_2) \in G_1 \,\circledS\, G_2 $.

\subsection{Reduced equations of motion}
\label{red_equ_motion}

Let $L:T(G_1\,\circledS\,G_2)\times V^*\rightarrow\mathbb{R}$ be a $G_1$-invariant Lagrangian under the affine action of $G_1$ on $T(G_1\,\circledS\,G_2) \times V ^\ast$ given by
\[
(g_1,\dot g_1,g_2,\dot g_2,a_0)\mapsto (hg_1,h\dot g_1,h\cdot g_2,h\cdot \dot g_2,(h,e)a_0+c(h,e)), \quad h\in G_1.
\]
The quotient space $\left(T(G_1\,\circledS\,G_2)\times V^*\right)/G_1$ is in a natural way isomorphic to the space $\mathfrak{g}_1\times TG_2\times V^*$, the associated projection being given by
\begin{align}
\label{metamorphosis_quotient}
&(g_1,\dot g_1,g_2,\dot g_2,a_0)\in T(G_1\,\circledS\, G_2)\times V^* \nonumber \\
& \qquad \qquad  \mapsto \left(g_1^{-1}\dot g_1,g_1^{-1}\cdot g_2,g_1^{-1}\cdot \dot g_2,\theta_{(g_1,e)^{-1}}a_0\right) \in \mathfrak{g}_1\times TG_2\times V^*.
\end{align}
This reduction map is a special case of the metamorphosis reduction described in \cite{HoGBRa2010}, with the exception that here one has an additional parameter $a_0$ that produces an advection equation in the  convective description.

In our particular case of two Lie groups, there is an additional isomorphism  that identifies the quotient space $\mathfrak{g}_1\times TG_2\times V^*$ with a simpler space, namely
\[
\mathfrak{g}_1\times TG_2\times V^*\rightarrow \mathfrak{g}_1\times G_2\times\mathfrak{g}_2\times V^*,\quad (\omega_1,p,\dot p,a)\mapsto \left(\omega_1,p,p^{-1}\dot p,\theta_{(e,p)^{-1}}a\right).
\]
By composing this isomorphism with the quotient map \eqref{metamorphosis_quotient}, we get the projection
\begin{align}\label{2_bouquet_quotient_map}
(g_1,\dot g_1,g_2,\dot g_2,a_0)
\mapsto &\left(g_1^{-1}\dot g_1,g_1^{-1}\cdot g_2,
g_1^{-1}\cdot \dot g_2,\theta_{(g_1,e)^{-1}}a_0\right)
\nonumber \\
\mapsto &\left(g_1^{-1}\dot g_1,g_1^{-1}\cdot g_2,(g_1^{-1}\cdot g_2)^{-1}(g_1^{-1}\cdot \dot g_2),\theta_{\left(e,g_1^{-1}\cdot g_2\right)^{-1}}\theta_{(g_1,e)^{-1}}a_0\right) \nonumber \\
&\quad  =\left(g_1^{-1}\dot g_1,g_1^{-1}\cdot g_2,g_1^{-1}\cdot( g_2^{-1}\dot g_2),\theta_{(g_1,g_2)^{-1}}a_0\right)
\end{align}
since $(g_1,e)\left(e,g_1^{-1}\cdot g_2\right)=(g_1,g_2)$. This quotient map will be shown to produce appropriate convective reduced variables in our applications. We denote by
\[
\ell:\mathfrak{g}_1\times G_2\times\mathfrak{g}_2\times V^*\rightarrow\mathbb{R}
\]
the associated reduced Lagrangian defined by 
\[
L(g_1,\dot g_1,g_2,\dot g_2,a_0)=\ell\left(g_1^{-1}\dot g_1,g_1^{-1}\cdot g_2,g_1^{-1}\cdot( g_2^{-1}\dot g_2),\theta_{(g_1,g_2)^{-1}}a_0\right)=:\ell(\omega_1,p,\omega_2,a),
\]
where we have introduced the reduced variables defined by
\begin{align*}
\omega_1&=g_1^{-1}\dot g_1\\
p&=g_1^{-1}\cdot g_2\\
\omega_2&=g_1^{-1}\cdot g_2^{-1}\dot g_2\\
a&=\theta_{(g_1,g_2)^{-1}}a_0=\theta_{(g_1,g_1\cdot p)^{-1}}a_0.
\end{align*}

In order to compute the reduced Euler-Lagrange equations,
we calculate the variations of these variables. A direct
computation yields
\begin{align*}
\delta\omega_1&=\dot\eta_1+[\omega_1,\eta_1],\qquad\eta_1=g_1^{-1}\delta g_1\\
\delta p&=p\eta_2-\eta_1\cdot p,\qquad\eta_2=g_1^{-1}\cdot g_2^{-1}\delta g_2\\
\delta\omega_2&=\dot\eta_2+[\omega_2,\eta_2]+\omega_1\cdot\eta_2-\eta_1\cdot\omega_2\\
\delta a&=-\eta_1 a-\eta_2 a-\partial_1c(\eta_1)-\partial_2c(\eta_2),
\end{align*}
where $\eta_i $ are
curves in $\mathfrak{g}_i$ vanishing at the endpoints and $\partial_i c :\mathfrak{g}_i \rightarrow V ^\ast$, $i =1,2$, are the partial derivatives of $c$, that is,
\[
\partial_1 c( \xi_1) : = \left.\frac{d}{dt}\right|_{t=0}
c( \exp (t \xi_1), e),\quad \text{and}\quad \partial_2 c( \xi_2) : = \left.\frac{d}{dt}\right|_{t=0}
c( e,\exp (t \xi_2)).
\]
The dots $(\,\cdot\,)$ appearing in the above expression denote the operations naturally induced by the action of $G_1$ on  $G _2$:
\begin{equation}\label{explanation_dot}
g_1\cdot\omega_2=\left.\frac{d}{dt}\right|_{t=0} g_1\cdot \exp (t \omega _2  ),\qquad \omega _1 \cdot \omega _2 =\left.\frac{d}{dt}\right|_{t=0} \exp (t \omega _1 )\cdot \omega _2,
\end{equation}
where $g_1\in G_1$, $ \omega _1 \in \mathfrak{ g} _1 $, $ \omega _2\in \mathfrak{ g} _2 $.
The formula for $\delta \omega _2 $ is proved in the following way
\begin{align*}
\delta \omega _2 = \delta (g_1^{-1}\cdot g_2^{-1}\dot g_2)&=-g_1^{-1}\delta g_1 g_1 ^{-1} \cdot g_2 ^{-1} \dot g_2-g_1 ^{-1} \cdot g_2 ^{-1} \delta g_2 g_2 ^{-1} \dot g_2+\partial_t (g_1 ^{-1} \cdot g_2 ^{-1} \delta g_2)\\
&\qquad +g_1 ^{-1} \dot  g_1 g_1 ^{-1} \cdot g_2 ^{-1} \delta g_2 +g_1 ^{-1}\cdot  g_2 ^{-1} \dot  g_2 g_2 ^{-1} \delta g_2\\
&=-\eta_1\cdot \omega _2 +\dot\eta_2
+\omega _1\cdot \eta_2+[\omega _2,\eta_2].
\end{align*}
To obtain the formula for $\delta a $ we recall that if a Lie group $G$ is represented on $V$, then
for $a=g^{-1}a_{0}+c(g^{-1}) \in V ^\ast$, we have $\delta a=-\eta a-\mathbf{d}c(\eta)$, where $\eta=g^{-1}\delta g$. In our case $G = G _1 \,\circledS\,G_2 $ so that taking $g = (g_1, g_1\cdot p) $ we get
\begin{align*}
g^{-1}\delta g&=(g_1,g_1\cdot p)^{-1}\delta (g_1,g_1\cdot p)=(g_1^{-1},p^{-1})(\delta g_1,\delta g_1\cdot p+g_1\cdot\delta p)\\
&=\left(g_1^{-1}\delta g_1,p^{-1}(g_1^{-1}\cdot (\delta g_1\cdot p+g_1\cdot(p\eta_2-\eta_1\cdot p)))\right)=:(\eta_1,\eta_2).
\end{align*}
Since
\[
(\eta_1,\eta_2)a=\left.\frac{d}{dt}\right|_{t=0}(\operatorname{exp}(t\eta_1),\operatorname{exp}(t\eta_2))a=\left.\frac{d}{dt}\right|_{t=0}(e,\operatorname{exp}(t\eta_2))(\operatorname{exp}(t\eta_1),e)a=\eta_1a+\eta_2a,
\]
the formula for $\delta a$ follows.

In order to write the reduced equations of motion, we need to define the following \emph{diamond operations}. The diamonds associated to the representation of $G_1\,\circledS\,G_2$ on $V^*$ are
\begin{equation}
\label{diamond_i}
\diamond_i:V\times V^*\rightarrow\mathfrak{g}_i^*,\quad \left\langle v\diamond_i a,\eta_i\right\rangle=-\left\langle \eta_ia,v\right\rangle=\left\langle a,\eta_iv\right\rangle.
\end{equation}
The diamond associated to the representation of $G_1$ on $\mathfrak{g}_2$ is
\[
\diamond_{12}:\mathfrak{g}_2\times\mathfrak{g}^*_2\rightarrow\mathfrak{g}_1^\ast,\quad \left\langle \xi_2\diamond_{12} \mu_2,\eta_1\right\rangle=\left\langle\mu_2,\eta_1\cdot \xi_2\right\rangle.
\]
Using the above constrained variations in 
\[
\delta\int_{t_1}^{t_2} \ell( \omega_1, p , \omega_2, a) dt  = 0,
\]
a long but straightforward computation finally yields the reduced Euler-Lagrange equations,
\begin{equation}\label{reduced_EL}
\left\{\begin{array}{l}
\vspace{0.2cm}\displaystyle\frac{d}{dt}\frac{\delta\ell}{\delta\omega_1}=\operatorname{ad}^*_{\omega_1}\frac{\delta\ell}{\delta\omega_1}-\mathbf{J}_{12}\left(\frac{\delta\ell}{\delta p}\right)-\omega_2\diamond_{12}\frac{\delta\ell}{\delta\omega_2}+\frac{\delta\ell}{\delta a}\diamond_1a-\partial_1c^T\left(\frac{\delta\ell}{\delta a}\right),\\
\displaystyle\frac{d}{dt}\frac{\delta\ell}{\delta\omega_2}=\operatorname{ad}^*_{\omega_2}\frac{\delta\ell}{\delta\omega_2}+p^{-1}\frac{\delta\ell}{\delta p}-\omega_1\cdot\frac{\delta\ell}{\delta\omega_2}+\frac{\delta\ell}{\delta a}\diamond_2a-\partial_2c^T\left(\frac{\delta\ell}{\delta a}\right),
\end{array}\right.
\end{equation}
to which we add the two equations
\begin{equation}
\label{kinematic_equations_abstract}
\dot p=p\omega_2-\omega_1\cdot p\qquad\text{and}\qquad \dot a+\omega_1a+\omega_2a+\partial_1c(\omega_1)+\partial_2c(\omega_2)=0.
\end{equation}
Here $\mathbf{J}_{12}: T ^\ast G_2 \rightarrow \mathfrak{g}^\ast_1 $ is the momentum map of the cotangent lifted action
of $G_1$ on $G_2$ given by
\[
\left\langle \mathbf{J}_{12}( \alpha_{g_2}), \xi_1 \right\rangle =\left\langle \alpha_{g_2},\xi_1\cdot g_2 \right\rangle,
\]
where
$\alpha_{g_2}\in T^\ast_{g_2} G_2$ and $\xi_1 \in \mathfrak{g}_1$.

By using the expression of the infinitesimal coadjoint 
action of the semidirect product $G_1\,\circledS\, G_2$, namely
\[
\operatorname{ad}_{( \omega_1, \omega_2)} ( \mu_1, \mu_2) =
\left(\operatorname{ad}_{ \omega_1} \mu_1 - \omega_2 \diamond_{12} \mu_2, \operatorname{ad}_{\omega_2} \mu_2 - \omega_1 \cdot \mu_2\right)
,
\]
for $\omega_1 \in \mathfrak{g}_1$, $\omega_2 \in \mathfrak{g}_2$, $\mu_1 \in \mathfrak{g}_1^\ast$, $\mu_2 \in \mathfrak{g}_2^\ast$,
equations \eqref{reduced_EL} can be written in a compact form as
\begin{align}
\label{compact_version_general}
&\frac{d}{dt}\left(\frac{\delta \ell}{\delta\omega_1},\frac{\delta \ell}{\delta\omega_1}\right) \nonumber \\
& \qquad =\operatorname{ad}^*_{(\omega_1,\omega_2)}\left(\frac{\delta \ell}{\delta\omega_1},\frac{\delta \ell}{\delta\omega_1}\right)+\frac{\delta\ell}{\delta a}\diamond a-\mathbf{d}c^T\left(\frac{\delta\ell}{\delta a}\right)+\left(-\mathbf{J}_{12}\left(\frac{\delta\ell}{\delta p}\right),p^{-1}\frac{\delta\ell}{\delta p}\right).
\end{align}

\paragraph{Conservative form.}
Given two curves $\mu(t)\in\mathfrak{g}^*$ and $g(t)\in G$, we have the formula
\begin{equation}\label{ad_Ad}
\frac{\partial}{\partial t}\operatorname{Ad}^*_{g(t)^{-1}}\mu(t)=\operatorname{Ad}^*_{g(t)^{-1}}\left(\frac{\partial}{\partial t}\mu(t)-\operatorname{ad}^*_{\xi(t)}\mu(t)\right)
\end{equation}
where $\xi(t)=g(t)^{-1}\dot g(t)\in \mathfrak{g}$. Suppose that the Lie group $G$ acts on a vector space $V$ by the representation $v\mapsto gv$. Consider the affine representation $a\mapsto ga+c(g)$ of $G$ on $V^*$, where $a\mapsto ga$ is the contragredient representation and $c$ a group one-cocycle. Then we have the relation
\begin{equation}\label{dc_Ad}
\mathbf{d}c^T\left(gv\right)=\operatorname{Ad}^*_{g^{-1}}\left(\mathbf{d}c^T\left(v\right)-v\diamond c(g^{-1})\right).
\end{equation}
By using these formulas and supposing that $a_0 = 0 $, equations \eqref{compact_version_general} can be rewritten in the form of a conservation law as
\begin{align}
\label{conservation_general}
&\frac{d}{dt}\operatorname{Ad}^*_{(g_1,g_2)^{-1}}\left(\frac{\delta\ell}{\delta\omega_1},\frac{\delta\ell}{\delta\omega_2}\right)+\mathbf{d}c^T\left((g_1,g_2)\frac{\delta \ell}{\delta a}\right) \nonumber \\
& \qquad =\operatorname{Ad}^*_{(g_1,g_2)^{-1}}\left(-\mathbf{J}_{12}\left(\frac{\delta\ell}{\delta p}\right),p^{-1}\frac{\delta\ell}{\delta p}\right).
\end{align}
This is easily seen since $a$ evolves as $a = \theta_{(g_1, 
g_2) ^{-1}} a_0 =(g_1, g_2)^{-1} a_0 + c((g_1, g_2)^{-1})$.

\subsection{Hamiltonian formulation and Poisson brackets}\label{Hamiltonian_formulation}

In this paragraph, we briefly describe the Hamiltonian side of the reduction process.

Let $H: T^*(G_1\,\circledS\, G_2)\times V^*\rightarrow\mathbb{R}$ be a $G_1$-invariant Hamiltonian. One can think of $H$ as being obtained from the Lagrangian $L$ by a Legendre transformation, the variable in $V^*$ being viewed as a parameter. On $T^*(G_1\,\circledS\, G_2)\times V^*$ we consider the Poisson structure given by the canonical symplectic form on $T^*(G_1\,\circledS\, G_2)$ and the trivial Poisson structure on $V^*$. In this way, Hamilton's equations for $H$ are equivalent to the canonical Hamilton equations for $H_{a_0}$ together with the equation $\dot a_0=0$. As on the Lagrangian side, we define $H_{a_0}(\alpha_{g_1},\alpha_{g_2}):=H(\alpha_{g_1},\alpha_{g_2},a_0)$.

As in \S\ref{red_equ_motion}, we consider the quotient map
\[
 T^*(G_1\,\circledS\, G_2)\times V^*\rightarrow \mathfrak{g}_1^*\times G_2\times\mathfrak{g}_2^*\times V^*,
\]
given by
\begin{equation}
\label{quotient_map}
\!\!\!
(\alpha_{g_1},\alpha_{g_2},a_0)\mapsto (\mu_1,p,\mu_2,a):=\left(g_1^{-1}\alpha_{g_1},g_1^{-1}\cdot g_2,g_1^{-1}\cdot (g_2^{-1}\alpha_{g_2}),\theta_{(g_1,g_2)^{-1}}(a_0)\right)
\end{equation}
which defines the reduced Hamiltonian $h=h(\mu_1,p,\mu_2,a)$ on
$\mathfrak{g}^\ast_1 \times G _2 \times \mathfrak{g}_2 \times V ^\ast$. If $H$ is associated to a Lagrangian $L$ by Legendre transformation, the reduced Hamiltonian $h$ can also be obtained from $\ell$ by the Legendre transformation, namely,
\begin{equation}
\label{Legendre_double_sd}
h(\mu_1,p,\mu_2,a)=\langle\mu_1,\omega_1\rangle+\langle\mu_2,\omega_2\rangle-\ell(\omega_1,p,\omega_2,a),\qquad \frac{\delta\ell}{\delta \omega_i}=\mu_i.
\end{equation}
To obtain the reduced Hamilton equations, one first needs to compute the Poisson bracket on $\mathfrak{g}_1^*\times G_2\times\mathfrak{g}_2^*\times V^*$ obtained by Poisson reduction (see e.g. \cite[\S10.7]{MaRa2002}). A 
direct computation yields
\begin{align*}
\{f,h\}=&-\left\langle \mu_1,\left[\frac{\delta f}{\delta\mu_1},\frac{\delta h}{\delta\mu_1}\right]\right\rangle-\left\langle \mu_2,\left[\frac{\delta f}{\delta\mu_2},\frac{\delta h}{\delta\mu_2}\right]\right\rangle-\left\langle\mu_2,\frac{\delta f}{\delta\mu_1}\cdot \frac{\delta h}{\delta\mu_2}-\frac{\delta h}{\delta\mu_1}\cdot \frac{\delta f}{\delta\mu_2}\right\rangle\\
& -\left\langle a,\frac{\delta f}{\delta \mu_1}\frac{\delta h}{\delta a}+\frac{\delta f}{\delta \mu_2}\frac{\delta h}{\delta a}-\frac{\delta h}{\delta \mu_1}\frac{\delta f}{\delta a}-\frac{\delta h}{\delta \mu_2}\frac{\delta f}{\delta a}\right\rangle\\
&+\left\langle \partial_1c\left(\frac{\delta f}{\delta \mu_1}\right)+\partial_2c\left(\frac{\delta f}{\delta \mu_2}\right),\frac{\delta h}{\delta a}\right\rangle-\left\langle \partial_1c\left(\frac{\delta h}{\delta \mu_1}\right)+\partial_2c\left(\frac{\delta h}{\delta \mu_2}\right),\frac{\delta f}{\delta a}\right\rangle\\
&+\left\langle \frac{\delta f}{\delta \mu_1},\mathbf{J}_{12}\left(\frac{\delta h}{\delta p}\right)\right\rangle+\left\langle\frac{\delta f}{\delta p},p\frac{\delta h}{\delta\mu_2}\right\rangle-\left\langle \frac{\delta h}{\delta \mu_1},\mathbf{J}_{12}\left(\frac{\delta h}{\delta p}\right)\right\rangle-\left\langle\frac{\delta h}{\delta p},p\frac{\delta f}{\delta\mu_2}\right\rangle.
\end{align*}
The first line represents the Lie-Poisson bracket of the dual 
of the semidirect product Lie algebra  
$\mathfrak{g}_1\,\circledS\,\mathfrak{g}_2$; when the 
second line is added, we get the Lie-Poisson bracket on 
the dual of the semidirect product $(\mathfrak{g}_1\,\circledS\,\mathfrak{g}_2)\,\circledS\, V$. The third line is due to the presence of the affine term. The last line is a new expression arising from the fact that reduction is carried out for the subgroup $G_1\subset G_1\,\circledS\,G_2$ and not the whole semidirect product. The Hamiltonian equations associated to this Poisson bracket are
\begin{equation}
\left\{\begin{array}{l}
\vspace{0.2cm}\displaystyle\frac{d}{dt}\mu_1=\operatorname{ad}^*_{\frac{\delta h}{\delta\mu_1}}\mu_1+\mathbf{J}_{12}\left(\frac{\delta h}{\delta p}\right)-\frac{\delta h}{\delta\mu_2}\diamond_{12}\mu_2-\frac{\delta h}{\delta a}\diamond_1a+\partial_1c^T\left(\frac{\delta h}{\delta a}\right),\\
\vspace{0.2cm}\displaystyle\frac{d}{dt}\mu_2=\operatorname{ad}^*_{\frac{\delta h}{\delta\mu_2}}\mu_2-p^{-1}\frac{\delta h}{\delta p}-\frac{\delta h}{\delta\mu_1}\cdot\mu_2-\frac{\delta h}{\delta a}\diamond_2a+\partial_2c^T\left(\frac{\delta\ell}{\delta a}\right),\\
\vspace{0.2cm}\displaystyle\frac{d}{dt}p=p\frac{\delta h}{\delta\mu_2}-\frac{\delta h}{\delta\mu_1}\cdot p,\\
\vspace{0.2cm}\displaystyle\frac{d}{dt}a=-\frac{\delta h}{\delta\mu_1}a-\frac{\delta h}{\delta\mu_2}a-\partial_1c\left(\frac{\delta h}{\delta\mu_1}\right)-\partial_2c\left(\frac{\delta h}{\delta\mu_2}\right).
\end{array}\right.
\end{equation}
If $h$ is obtained from a Lagrangian $\ell $ by the Legendre transformation \eqref{Legendre_double_sd}, these equations can be obtained by substituting the relations
\[
\frac{\delta h}{\delta\mu_i}=\omega_i,\qquad\frac{\delta h}{\delta p}=-\frac{\delta \ell}{\delta p},\qquad\frac{\delta h}{\delta a}=-\frac{\delta \ell}{\delta a},
\]
into the reduced Euler-Lagrange equations \eqref{reduced_EL}, \eqref{kinematic_equations_abstract}.

\subsection{Affine reduction at fixed parameter}
\label{sec_fixed_par}

In applications, one often has the situation that the Lagrangian $L_{a_0}:T(G_1 \,\circledS\,G_2) \rightarrow \mathbb{R}$ is known only for a fixed value $a_0 \in V^\ast$. In this case, the theory developed so far does not apply,
since $\ell$ is not the reduction of a $G_1$-invariant function on $T(G_1 \,\circledS\,G_2) \times V^\ast$. It
turns out that the theory in Section \ref{red_equ_motion} is applicable, however, if we assume that
$L_{a_0} $ is $(G_1)^c_{a_0}$-invariant, where
\[
(G_1)^c_{a_0}: = \{ g_1 \in G_1 \mid \theta_{(g_1, e)} a_0 
= a_0\}.
\]
It also happens in applications (this is the case for the bouquets) that the reduced Lagrangian $\ell$ cannot be explicitly expressed
as a function of the variables $\xi_1, p, \xi_2, a$ but it has the form
$\ell(\xi_1, p, \xi_2, a, g_1) $, where $g_1 \in G_1$ is such that $c(g_1 ^{-1}, p ^{-1}) = a$.

 If we assume that 
$\ell$ is $(G_1)^c_{a_0}$-invariant, that is, $\ell(\xi_1, p, \xi_2, a, hg_1) = \ell(\xi_1, p, \xi_2, a, g)$ for all $h \in (G_1)^c_{a_0}$, then $\ell(\xi_1, p, \xi_2, a, g)$ is a well defined function of $(\xi_1, p, \xi_2, a)$. Indeed,
if $c(g_1 ^{-1}, p^{-1}) = c(h_1^{-1}, p^{-1})$, then
$\theta_{(g_1 ^{-1}, p^{-1})} 0 = \theta_{(h_1^{-1}, p^{-1})}0$ which is equivalent to 
\[
(h_1^{-1}, p^{-1})^{-1}
(g_1 ^{-1}, p^{-1}) = (h_1 g_1 ^{-1}, e) \in (G_1 \,\circledS\, G_2)^c_0,
\]
that is, $h_1g_1 ^{-1} \in (G_1)^c_0 $. In this case, equations 
\eqref{reduced_EL} are still valid, where one
computes the functional derivatives as if $g_1$ were expressed
explicitly in terms of $\xi_1, p, \xi_2, a$. Although this may
not actually be possible, the derivatives may still have explicit expressions
in many applications. We refer for details to 
\cite{ElGBHoPuRa2010}.

\subsection{Generalization to $N$ groups} 
\label{sec:generalization} 

In this section we introduce iterated semidirect products 
that will be crucial in the description of the dynamics
of multi-bouquets. This is motivated by the fact that, as we have mentioned earlier, 
the rotation of a bouquet at the $k$th level induces, via actions of semidirect products, rotation of all subsequent levels of bouquets, and thus the distance between the charges is affected correspondingly. The semidirect product rotation is a mathematical property and it is applicable to any dendronized polymer with a tree-like structure of the bouquets. This representation is independent of the details of the polymer, its chemical composition, \emph{etc.}, as it represents the purely topological aspects of the polymer's geometry. 

Note that,  in principle, our construction fails should some of the bouquets close at a subsequent stage (so the dendrimes become a graph and not a tree). Such graph-like molecules have not been synthesized yet. There are also substantial mathematical challenges in dealing with graph-like structures as opposed to tree-like structures. In this paper we will
study exclusively molecules with tree-like structures.

\paragraph{Semidirect product structure.} We shall present the construction of the configuration space of $N $-level bouquets, also called $N $-bouquets. This will be an iterated semidirect product of $N $ Lie groups modeling the tree structure of a multi-bouquet.

Let $G _1, \ldots, G_N $
be Lie groups. Assume that $G_{N-1}$ acts on $G_N$ by group homomorphisms and form the associated semidirect product group $G_{N-1}\,\circledS\, G_N$. Then suppose that the group $G_{N-2}$ acts on $G_{N-1}\,\circledS\, G_N$ by group homomorphisms, that is,
\[
g_{N-2}\cdot \left((g^1_{N-1},g^1_{N-2})(g^2_{N-1},g^2_{N-2})\right)=\left(g_{N-2}\cdot (g^1_{N-1},g^1_{N-2})\right) \left(g_{N-2}\cdot (g^2_{N-1},g^2_{N-2})\right)
\]
and form the semidirect product $G_{N-2} \,\circledS\,(
G_{N-1}\,\circledS\, G_N)$. Proceed inductively and form
the iterated semidirect product group 
\begin{equation}
\label{N_semidirect_product}
G_1\,\circledS\,(G_2\,\circledS\,(G_3\,\circledS\,(\cdots\,\circledS\, G_N)\cdots ))\ni (g_1,g_2,\ldots,g_N).
\end{equation}

For example, suppose that $G $ acts by group homomorphisms
on two other groups $H $ and $K $. Then 
\[
g\cdot (h,k):=(g\cdot h,g\cdot k), \quad g \in G, \; h \in H, \; k \in K,
\]
is an action of $G $ on $H \,\circledS\,K $ by group homomorphisms, provided the following compatibility condition holds:
\[
g\cdot (h\cdot k)=(g\cdot h)\cdot (g\cdot k).
\]

A simple example of such a chain of semidirect products is obtained by considering the same group $G$, acting on itself by conjugation: $G\,\circledS\,(G\,\circledS\,(G\,\circledS\,(\cdots\,\circledS\, G)\cdots))$.

Returning to the general case, the group multiplication reads
\[
(g_1,g_2,...,g_N)(h_1,h_2,...,h_N)=\big(g_1h_1,g_2(g_1\cdot h_2),g_3(g_2\cdot (g_1\cdot h_3)),...,g_N(g_{N-1}\cdot ...\cdot g_2\cdot (g_1\cdot h_N))\big)
\]
and the inverse is
\[
(g_1,g_2,...,g_N)^{-1}=\big(g_1^{-1},g_1^{-1}\cdot g_2^{-1},g_1^{-1}\cdot (g_2^{-1}\cdot g_3^{-1}),...,g_1^{-1}\cdot (g_2^{-1}\cdot ...\cdot(g_{N-1}^{-1}\cdot g_N^{-1}))\big).
\]

\paragraph{Lagrangian reduction.} Suppose that the semidirect product \eqref{N_semidirect_product} acts on the dual $V^*$  of a
vector space $V $ by an \emph{affine representation}
\[
a\mapsto\theta_{(g_1,...,g_N)}a= (g_1,..,g_N)a +
c(g_1,...,g_N), \quad a \in V^\ast.
\]
This includes all parameters given by the specific
physical situation in the $N$-bouquet.\smallskip

Let $L:T\left(G_1\,\circledS\,(G_2\,\circledS\,(G_3\,\circledS\,(\cdots\,\circledS\, G_N)\cdots))\right)\times V^*\rightarrow\mathbb{R}$ be a $G_1$-invariant Lagrangian under the affine action of $G_1$ on $T\left(G_1\,\circledS\,(G_2\,\circledS\,(G_3\,\circledS\,(\cdots\,\circledS\, G_N)\cdots))\right) \times V ^\ast$ given by
\[
(g_1,\dot g_1,g_2,\dot g_2,...,g_N,\dot g_N,a_0)\mapsto 
\left(hg_1,h\dot g_1,h\cdot g_2,h\cdot \dot g_2,...,h\cdot g_N,h\cdot \dot g_N,\theta_{(h,e,..,e)}a_0\right), 
\]
$h\in G_1$. We consider the quotient map
\[
T\left(G_1\,\circledS\,(G_2\,\circledS\,(G_3\,\circledS\,(\cdots\,\circledS\, G_N)\cdots))\right)\times V^*\rightarrow\mathfrak{g}_1\times G_2\times\mathfrak{g}_2\times...\times G_N\times\mathfrak{g}_N
\]
given by
\begin{align*}
&(g_1,\dot g_1,g_2,\dot g_2,g_3,\dot g_3,...,g_N,\dot g_N,a_0)\mapsto(\omega_1,p_2,\omega_2,p_3,\omega_3,...,p_N,\omega_N,a)\\
&\qquad:=\left(g_1^{-1}\dot g_1,g_1^{-1}\cdot g_2,g_1^{-1}\cdot g_2^{-1}\dot g_2,g_1^{-1}\cdot g_2^{-1}\cdot g_3,g_1^{-1}\cdot g_2^{-1}\cdot g_3^{-1}\dot g_3,...,\theta_{(g_1,g_2,...,g_N)^{-1}}a_0\right).
\end{align*}
We thus obtain the convective variables
\[
\omega_i=g_1^{-1}\cdot g_2^{-1}\cdot ... \cdot g_i^{-1}\dot g_i\qquad\text{and}\qquad p_i=g_1^{-1}\cdot g_2^{-1}\cdot ... \cdot g_{i-1}^{-1}\cdot g_i.
\]
Using the semidirect product multiplication, we can write
\[
(\omega_1,\omega_2,...,\omega_N)=(g_1,g_2,...,g_N)^{-1}(\dot g_1,\dot g_2,...,\dot g_N).
\]
Note that the group $G_1\,\circledS\,(...\,\circledS\,G_{N-1})$ naturally acts on the group $G_2\,\circledS\,(...\,\circledS\,G_N)$. The action is given by the same expression as the group multiplication on the semidirect product:
\begin{equation}\label{induced_action}
\!\!\!\!\!
(g_1,...,g_{N-1})\bullet (h_2,...,h_N):=\left(g_1\cdot h_2,g_2\cdot (g_1\cdot h_3),...,g_{N-1}\cdot(g_{N-2}\cdot...\cdot g_1\cdot h_N)\right).
\end{equation}
Using this action, the variables $p_i$ can be obtained by the formula
\[
(p_2,...,p_N)=(g_1,...,g_{N-1})^{-1}\bullet (g_2,...,g_N).
\]

\paragraph{Reduced Euler-Lagrange equations.} A straightforward computation shows that the constrained variations are
\begin{equation}
\delta\omega_i=\frac{\partial}{\partial t}\eta_i+[\omega_i,\eta_i]
+\sum_{k=1}^{i-1}
(\omega_k\cdot\eta_i-\eta_k\cdot\omega_i)
,\quad \eta_i:=g_1^{-1}\cdot (g_2^{-1}\cdot (...\cdot g_i^{-1}\delta g_i))
\label{deltaomegai} 
\end{equation}
and
\begin{equation}
\delta p_i=p_i\eta_i-\sum_{k=1}^{i-1}\eta_k\cdot p_i.
\label{deltapi} 
\end{equation}
where $\eta_i$ are curves in $\mathfrak{g}_i$ vanishing at the endpoints, $i=1, \ldots, N $. The dots in the above expression denote the operations naturally induced from the action of $G_k$ on $G _i $, $k<i$, as in \eqref{explanation_dot}.

For example, the formula for $\delta \omega _i$ is proved in the following way
\begin{align*}
\delta \omega _i &= \delta (g_1^{-1}\cdot g_2^{-1} \cdot .... \cdot g_i ^{-1} \dot g_i)\\
&=- \left( g_1^{-1}\delta g_1 g_1 ^{-1} \cdot g_2 ^{-1}  \cdot .... \cdot  g_i ^{-1} \dot g_i \right) - \left( g_1 ^{-1} \cdot g_2 ^{-1} \delta g_2 g_2 ^{-1} \cdot ... \cdot  g _i ^{-1} \dot{ g}_i \right) \\
&\qquad-... - \left( g _1 ^{-1} \cdot g _2 ^{-1} \cdot ... \cdot g _i ^{-1} \delta g _i g _i ^{-1} \dot{ g} _i \right) + \left( g _1 ^{-1} \cdot g _2 ^{-1} \cdot ... \cdot g _i ^{-1} \delta \dot{ g} _i  \right) \\
&=-\sum _{k=1}^{i-1} \eta  _k \cdot \omega _i - \eta _i \omega _i +\frac{\partial  }{\partial t }  \left( g _1 ^{-1} \cdot g _2 ^{-1} \cdot ... \cdot g _i ^{-1} \delta g _i  \right) +\left( g_1^{-1} \dot{ g}_1 g_1 ^{-1} \cdot g_2 ^{-1}  \cdot .... \cdot  g_i ^{-1} \delta  g_i \right) \\
&\qquad +\left( g_1 ^{-1} \cdot g_2 ^{-1}\dot{g}_2 g_2 ^{-1} \cdot ... \cdot  g _i ^{-1} \delta g_i \right) +... + \left( g _1 ^{-1} \cdot g _2 ^{-1} \cdot ... \cdot g _i ^{-1} \dot{ g} _i g _i ^{-1} \delta { g} _i \right)\\
&=-\sum _{k=1}^{i-1} \eta  _k \cdot \omega _i - \eta _i \omega _i + \frac{\partial  }{\partial  t}  \eta _i + \sum _{k=1}^{i-1} \omega _k \cdot \eta _i + \omega _i \eta _i \\
&=\frac{\partial}{\partial t}\eta_i+[\omega_i,\eta_i]
+\sum_{k=1}^{i-1}
(\omega_k\cdot\eta_i-\eta_k\cdot\omega_i)
\end{align*}

Using these variations in
\[
\delta\int_{t_0}^{t_1}\ell(\omega_1,p_2,\omega_2,p_3,\omega_3,...,p_N,\omega_N,a)dt=0
\]
yields the $N$ equations
\begin{align}\label{equations_N_bouquets}
\frac{d}{dt}\frac{\delta \ell}{\delta\omega_i}
=
\operatorname{ad}^*_{\omega_i}\frac{\delta\ell}{\delta\omega_i}&
-
\sum_{j=1}^{i-1}
\omega_j\cdot\frac{\delta \ell}{\delta\omega_i}
+
p_i^{-1}\frac{\delta\ell}{\delta p_i}
-
\sum_{k=i+1}^N\left(\omega_k\diamond_{ik}\frac{\delta\ell}{\delta\omega_k}+\mathbf{J}_{ik}\left(\frac{\delta\ell}{\delta p_k}\right)\right)\nonumber\\
&\qquad+\frac{\delta\ell}{\delta a}\diamond_i a-\partial_ic^T\left(\frac{\delta\ell}{\delta a}\right),\quad i=1,...,N,
\end{align}
where $\diamond_i: V \times V ^\ast \rightarrow 
\mathfrak{g}^\ast_i$ is given by \eqref{diamond_i}, and for $i< k$
\[
\diamond_{ik}: \mathfrak{g}_k\times \mathfrak{g}^\ast_k \rightarrow \mathfrak{g}^\ast_i, \quad \left\langle \xi_k \diamond _{ik} \mu_k, \eta_i \right\rangle : = \left\langle \mu_k, \eta_i\cdot \xi_k \right\rangle, \quad \xi_k \in \mathfrak{g}_k, \; \mu_k\in \mathfrak{g}^\ast_k,\; \eta_i \in \mathfrak{g}_i,
\]
and $\mathbf{J}_{ik} : T ^\ast G _k \rightarrow \mathfrak{g}^\ast_i$ is the momentum map of the cotangent lifted action
of $G_i$ on $G _k$.

Among these $N$ equations, note that the first and the last of them are simpler, since one of the terms involving the summation is absent. Note also that when $i=1$, the term $p_i^{-1}\frac{\delta\ell}{\delta p_i}$ is absent. (The $p_i$'s start with $p_2$.)

Using the infinitesimal coadjoint action of the semidirect product allows these equations to be rewritten as
\begin{align*}
\frac{d}{dt}\left(\frac{\delta\ell}{\delta \omega_1},...,\frac{\delta\ell}{\delta \omega_N}\right)=&\operatorname{ad}^*_{(\omega_1,...,\omega_N)}\left(\frac{\delta\ell}{\delta \omega_1},...,\frac{\delta\ell}{\delta \omega_N}\right)+\frac{\delta\ell}{\delta a}\diamond a-\mathbf{d}c^T\left(\frac{\delta\ell}{\delta a}\right)\\
&\qquad +\left(0,p_2^{-1}\frac{\delta\ell}{\delta p_2},...,p_N^{-1}\frac{\delta\ell}{\delta p_N}\right)-\left(\boldsymbol{\mathcal{J}}\left(\frac{\delta\ell}{\delta p}\right),0\right),
\end{align*}
where $\diamond : V ^\ast \times V \rightarrow [\mathfrak{g}_1 \,\circledS\,( \mathfrak{g}_2\,\circledS\,( \cdots \,\circledS\, \mathfrak{g}_N) \cdots )]^\ast $ is associated to the action of the whole semidirect product on $V ^\ast$ and
\[
\boldsymbol{\mathcal{J}}:T^*(G_2\,\circledS\,(\cdots\,\circledS\, G_N))\rightarrow (\mathfrak{g}_1\,\circledS\,(\cdots\,\circledS\,\mathfrak{g}_{N-1}))^*
\]
is the cotangent bundle momentum map associated to the action of $G_2\,\circledS\,(\cdots\,\circledS\, G_N)$ on $G_1\,\circledS\,(\cdots\,\circledS\, G_{N-1})$ defined in \eqref{induced_action}. It is given by
\[
\boldsymbol{\mathcal{J}}(\alpha_{g_2},...,\alpha_{g_N})=\left(\sum_{k=2}^N\mathbf{J}_{1k}(\alpha_{g_k}),...,\sum_{k=i+1}^N\mathbf{J}_{ik}(\alpha_{g_k}),...,\mathbf{J}_{N-1\, N}(\alpha_{g_N})\right)
.\]

\paragraph{Conservative form.} If $a_0 = 0$, equations \eqref{ad_Ad} and \eqref{dc_Ad} may be used to rewrite the equations of motion \eqref{equations_N_bouquets} equivalently in a conservative form as
\begin{align}\label{conservation_law}
&\frac{d}{dt}\operatorname{Ad}^*_{(g_1,...,g_N)^{-1}}\left(\frac{\delta\ell}{\delta\omega_1},...,\frac{\delta\ell}{\delta\omega_N}\right)+\mathbf{d}c^T\left((g_1,...,g_N)\frac{\delta \ell}{\delta a}\right)\nonumber\\
&\qquad\qquad=\operatorname{Ad}^*_{(g_1,...,g_N)^{-1}}\left(\left(0,p_2^{-1}\frac{\delta\ell}{\delta p_2},...,p_N^{-1}\frac{\delta\ell}{\delta p_N}\right)-\left(\boldsymbol{\mathcal{J}}\left(\frac{\delta\ell}{\delta p}\right),0\right)\right).
\end{align}

\section{Application to multi-bouquets}

\subsection{Dynamics of two-bouquets}

In order to apply these ideas to the dendronized polymers, we must parametrize the position of each charge. For the two-level polymers presented in Figure~\ref{fig:multi-bouquet}, the position of the base line is denoted by $\br(s,t)$. 
Then, at every point $s$, the rotation of the first level of the rigid bouquet (colored in red) is measured by $\Lambda_1 \in SO(3)$. In the exact geometric rod approach \cite{SiMaKr1988}, $\Lambda_1$ is measured with respect to a fixed coordinate system.  In order to complete the description of the system, the rotation of the  second level of bouquets (marked green) needs to be taken into account. In the exact geometric rod theory, the rotation of the second bouquet level $\Lambda_2 \in SO(3)$ would be measured with respect to the same fixed coordinate system, rather than with respect to the first bouquet. However, we shall depart from this convention and measure the orientation of each successive charge bouquet with respect to a frame attached to the previous bouquet. The latter convention will lead to a \emph{semidirect product action by homomorphisms}. 

The rotations of the base bouquet $\Lambda_1$ depend on the point on the base curve $s$ and, of course, time $t$. The 
rotation of the second bouquet level $\Lambda_2$ also depends on the numbered vertex to which it is attached. We call this vertex $k_1$ for $s$ and $m_1$ for $s'$. The position of a given charge depends on the number of the charge in the 
second bouquet (called $k_2$ for $s$ and $m_2$ for $s'$), on the orientations $\Lambda_1$ and $\Lambda_2^{k_1}$ and on the position of the base $\br(s)$. Thus, $ \Lambda _2$ is measured with respect to a frame attached to the first bouquet and we find the formulas 
\begin{align*}
\boldsymbol{c}_{k_1k_2}(s)&=\br(s)+\Lambda_1(s)\boldeta_{k_1}(s)+\Lambda_2^{k_1}(s)\Lambda_1(s)\boldeta_{k_2}(s)
,
\\
\boldsymbol{c}_{m_1m_2}(s')&=\br(s')+\Lambda_1(s')\boldeta_{m_1}(s')+\Lambda_2^{m_1}(s')\Lambda_1(s')\boldeta_{m_2}(s')
,
\end{align*}
for the position of charges on a two-bouquet anchored along the centerline at positions $\br(s)$ and $\br(s')$. Note that the orientation $\Lambda_i$ refers to the frame in which a charge bouquet at the $i$th level is rigidly positioned and not to the orientations of individual components of that charge bouquet. 

If more levels of the bouquets were present, we would need to introduce orientations $\Lambda_3$, $\Lambda_4$, \ldots  and corresponding labels $k_3, k_4 , \ldots$ at $s$, and $m_3, m_4, \ldots$ at $s'$. As we shall explain below, this notation is rather cumbersome, especially in the case of $N$-bouquets.  
We shall concentrate mainly on two-level bouquets here, as this case illustrates most of the mathematical concepts and is the necessary first step in applications. Formulas for $N$-level dendrimers will also be derived and discussed later in enough detail to make the general pattern of the equations evident. 

Let us explain how the group $SE(3) \,\circledS\,SO(3)$ appears already in the study of the two-level bouquet problem. We first consider the case of a \textit{single} multi-bouquet whose kinematic description coincides with that of a chain of rigid bodies.
\begin{figure}[ht]
\begin{center}
%\captionsetup{width=0.8 \textwidth}
\includegraphics[width=0.3\textwidth,angle=90]
{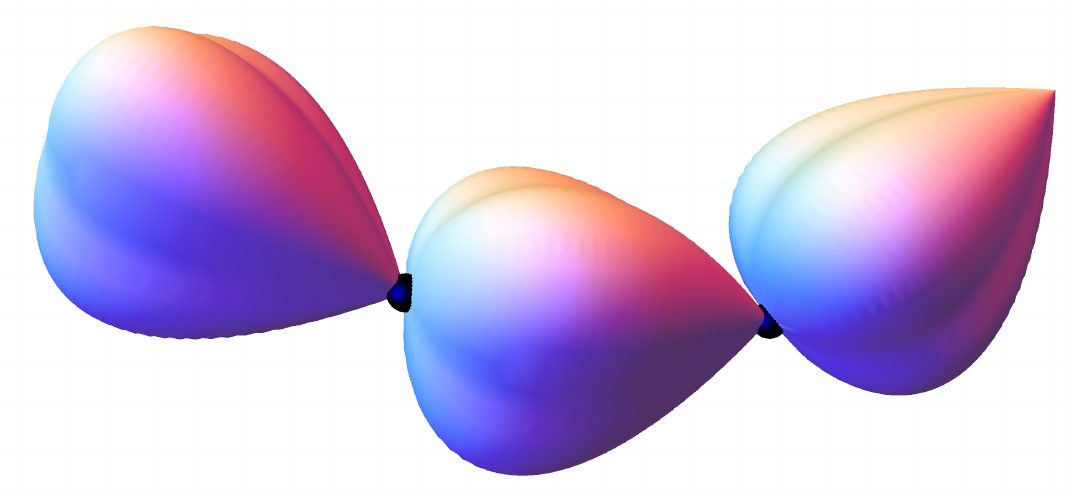}
\end{center}
\setlength{\captionmargin}{45pt}
\caption{\small{A figure illustrating a chain of coupled rigid bodies connected through a ball joint (black spheres). 
}}
\label{fig:rigid-bodies}
\end{figure}

\paragraph{Motivation for the semidirect product:  coupled rigid bodies.} Consider the kinematics of two rigid bodies coupled with a hinge joint, as in \cite{GrKrMa1988}. We denote by $\mathcal{B} _1$ and $\mathcal{B} _2$ the reference configurations of the bodies. Let $P_0$ be a preferred point (for example the center of mass) in the reference configuration  $ \mathcal{B} _1$. Let $P_1$ be the (fixed) hinge point in the reference configuration of the first body and $P_2$ be the (fixed) hinge point in the configuration of the second body. Let $\boldsymbol{ \eta }$ denote the vector from $P_0$ to $P_1$ in $\mathcal{B} _1 $.
Fix an inertial frame and denote by $ \boldsymbol{ r}(t)$ the current position of the point $P_0$. The configuration of the first body is determined by an element $\Lambda _1 (t)\in SO(3)$. We denote by $\Lambda_2(t)\in SO(3)$, the orientation of the second body \textit{relative to the first body}. The motion of a point in the first body is thus given by
\begin{equation} \label{motion_2_bodies_first}
x(t)= \boldsymbol{r}(t)+ \Lambda _1 (t)X,\qquad 
X\in \mathcal{B} _1 
\end{equation} 
and in the second by
\begin{equation} \label{motion_2_bodies_second}
x(t)= \boldsymbol{r}(t)+ \Lambda _1 (t)\boldsymbol{\eta }+\Lambda _2(t) \Lambda _1(t)Y,\qquad Y\in \mathcal{B} _2.
\end{equation} 
The configuration space of the system is $\mathbb{R}  ^3\times SO(3)\times SO(3)\ni ( \boldsymbol{ r}, \Lambda _1 , \Lambda _2 )$. In order to show how the semidirect product $SE(3) \,\circledS\,SO(3)$ arises, we consider the kinetic energy of the system
\begin{align*}
&K\left(\Lambda_1, \dot{\Lambda}_1, \boldsymbol{r}, 
\dot{\boldsymbol{r}}, \Lambda_2 , \dot{\Lambda}_2\right)
=\frac{1}{2}\int_{\mathcal{B}_1}\left|\dot{ x}(t)\right|^2 
\mbox{d}^3X +
\frac{1}{2} \int_{\mathcal{B}_2 }\left|\dot{y}(t)\right|^2 \mbox{d}^3Y\\
&\qquad \qquad =
\frac{1}{2} \int_{ \mathcal{B} _1 }\left |\dot{\boldsymbol{ r}}+\dot{ \Lambda} _1X\right |^2\mbox{d}^3X+\frac{1}{2} \int_{ \mathcal{B} _2 }\left |\dot{\boldsymbol{ r}}+\dot{ \Lambda} _1 \boldsymbol{ \eta }+\dot{ \Lambda} _2  \Lambda _1Y + \Lambda _2 \dot{ \Lambda} _1 Y\right |^2\mbox{d}^3Y,
\end{align*}  
where $\mbox{d}^3X$ and $\mbox{d}^3Y$ denote the mass densities of the corresponding bodies. One observes that the kinetic energy $K$ is invariant under the cotangent lift of the $SE(3)$ action given by
\[
( \Lambda _1, \boldsymbol{ r}, \Lambda _2 ) \mapsto \left( h \Lambda _1 ,h \boldsymbol{ r}+ \boldsymbol{ v}, h \cdot \Lambda _2 \right),
\]
where $ h \cdot \Lambda$ is the action by \textit{homomorphisms} defined by conjugation $h \cdot \Lambda := h \Lambda h ^{-1}$. The configuration space is therefore naturally given by $SE(3)\,\circledS\, SO(3)$.
By $SE(3)$ invariance, one can write the kinetic energy in terms of the reduced variables
\[
\omega _1 = \Lambda _1 ^{-1} \dot{ \Lambda }_1 ,\quad \boldsymbol{ \gamma }= \Lambda _1 ^{-1} \dot{ \boldsymbol{ r}},\quad p= \Lambda _1 ^{-1} \cdot \Lambda _2 ,\quad \omega _2 =p ^{-1} ( \Lambda _1 ^{-1} \cdot \dot{ \Lambda } _2) =\Lambda _1 ^{-1} \Lambda _2 ^{-1} \dot{ \Lambda }_2 \Lambda _1 ^{-1}
\]
on $T(SE(3) \,\circledS\, SO(3))/SE(3)\simeq \mathfrak{ se}(3)\times SO(3)\times \mathfrak{so}(3) $. If the Lagrangian of the system is only $SO(3)$ invariant one has the additional reduced variable $\boldsymbol{ \rho }= \Lambda _1^{-1}  \boldsymbol{ r}$; such a situation arises in the case of multi-bouquets. In order to obtain the equations of motion, we apply the framework described in Section \ref{sec_geometric_setting} with $G_1=SE(3)$, $G_2=SO(3)$, and the cocycle is absent. In the case of $SE(3)$-invariance, the abstract system \eqref{reduced_EL} gives
\begin{equation}
\left\{
\begin{aligned}
&\Big(\partial_t + \bom_1\times \Big)\dede{\ell}{\bom_1} 
+\bgam\times\dede{\ell}{\bgam} 
+\bom_2\times\dede{\ell}{\bom_2}+ \frac{\delta \ell}{\delta p} p^{-1}
- p^{-1}\frac{\delta \ell}{\delta p}  =0 \,,\\
&\Big( \prt_t + \bom_1  \times\Big)
 \dede{\ell}{\bgam}
  =0 \,,\\
&\Big( \prt_t + \bom_2\times\Big)\dede{\ell}{\bom_2} +
\bom_1\times\frac{\delta \ell}{\delta\bom_2}
- p^{-1}\frac{\delta \ell}{\delta p}  =0.
\end{aligned}
\right.
\end{equation}
The variables used here differ from those used in \cite{GrKrMa1988} where the configurations $A_1, A_2\in SO(3)$ of the bodies are given relative to the \emph{same} inertial frame. That is, we have the relations
\[
A_1=\Lambda _1 \quad A_2=\Lambda _2 \Lambda _1
\]
and \eqref{motion_2_bodies_first}, \eqref{motion_2_bodies_second} read
\begin{equation} \label{motion_2_bodies_GrKrMa}
x(t)= \boldsymbol{ r}(t)+ A_1 (t)X,\;\; X\in \mathcal{B} _1 \quad\text{resp.} \quad x(t)= \boldsymbol{ r}(t)+ A _1 (t)\boldsymbol{ \eta }+A_2(t)Y,\;\; Y\in \mathcal{B} _2.
\end{equation} 
Note that when $( \Lambda _1, \Lambda _2 )\mapsto (h \Lambda _1 , h \Lambda _2 h ^{-1})$ then $(A _1 , A _2 ) \mapsto (h A_1 , h A _2 )$, consistently with the symmetry considered in \cite{GrKrMa1988}. 
The reduced variables $\Omega_1=A_1 ^{-1} \dot{ A}_1$, $\Omega _2=A _2 ^{-1} \dot{ A} _2 $, and $A=A _1 ^{-1} A _2 $ used in \cite{GrKrMa1988} are recovered from our variables $ \omega _1$, $\omega _2$, $p$ via the simple relations
\begin{align*}
\Omega _1 &=\omega _1\,,\\
\Omega _2 &=A _2 ^{-1} \dot{ A} _2= (\Lambda _2 \Lambda _1 ) ^{-1} \left( \dot{ \Lambda } _2 \Lambda _1 + \Lambda _2 \dot{ \Lambda } _1 \right) = \omega _1 + \omega _2 \,,\\
A&=A _1 ^{-1} A _2 = \Lambda _1 ^{-1} \Lambda _2 \Lambda _1=p.
\end{align*}
On the Hamiltonian side, one has a Poisson diffeomorphism between the two representations.

\begin{remark}\rm
In \cite{GrKrMa1988}, the coupled system is actually described slightly differently than in equation \eqref{motion_2_bodies_GrKrMa}. Namely, a point in the first body, $X_1\in \mathcal{B} _1$, evolves as
$\boldsymbol{w}(t)+A_1(t)X_1$, while a point in the second body, $X_2\in \mathcal{B} _2$, evolves as $\boldsymbol{w}(t)+A_2(t)X_2$, where $ \boldsymbol{ w}(t)$ is the current position of the hinge between the two bodies. In order to simplify the comparison  with two-bouquets, we have changed the description slightly by using \eqref{motion_2_bodies_GrKrMa} instead, since this approach generalizes easily to a chain of $N$ rigid bodies, as we discuss below.\quad $\blacklozenge$
\end{remark}

In the case of a chain of $N$ coupled rigid bodies, the current position of a point in the $i$th rigid body is
\[
x(t)= \boldsymbol{ r}(t)+ \Lambda _1 (t)\boldsymbol{ \eta}_1+ \Lambda _2(t) \Lambda _1(t) \boldsymbol{ \eta }_2+...+\Lambda _i(t)...\Lambda _1(t)X_i,\;\; X_i\in \mathcal{B} _i,
\]
where $\mathcal{B}_i$ is the reference configuration of the $i$th body and $ \boldsymbol{ \eta }_i$ is the vector between the two hinges in the $i$th rigid body. Therefore, the kinetic energy is given by
\[
K( \Lambda _1, \dot{ \Lambda }_1, \boldsymbol{ r}, \dot{ \boldsymbol{ r}}, \Lambda _2, \dot{ \Lambda }_2,..., \Lambda _N, \dot{ \Lambda }_N)= \frac{1}{2} \sum_{i=1}^N\int_{ \mathcal{B} _i}\left| \dot{ \boldsymbol{ r}}+ \dot{\Lambda} _1 \boldsymbol{ \eta }_1+...+ \partial _t \left(\Lambda _i...\Lambda _1\right) X_i\right|^2\mbox{d}^3X_i,
\]
and one observes that $K$ is $SE(3)$-invariant under the action
\[
( \Lambda _1, \boldsymbol{ r}, \Lambda _2,..., \Lambda _N ) \mapsto \left( h \Lambda _1 ,h \boldsymbol{ r}+ \boldsymbol{ v}, h \cdot \Lambda _2,..., h \cdot \Lambda _N \right),
\]
in which the group $SE(3)$ acts by homomorphisms on the other $SO(3)$ groups. The configuration space is therefore naturally given by the iterated semidirect product
\[
SE(3)\,\circledS\,(SO(3)\,\circledS\,(SO(3)\,\circledS\,(\cdots\,\circledS\, SO(3))\cdots)).
\]

\begin{remark}\rm 
Another difference between coupled rigid bodies and our system of dendrite polymers is that each rigid body in the chain (say, at the position $i$) is completely characterized by its orientation matrix $\Lambda_i$. For a dendrite, 
 each branch of the bouquet at each level may have a different orientation.
 For example, suppose that the first level rigid bouquet consists of $n_1$ branches $\boldsymbol{\eta}_{k_1}$, $k_1=1,...,n_1$ whose orientation at $s$ is given by $ \Lambda _1 (s)$. At each extremity of these branches, there is a second level rigid bouquet given by $n_2$ branches $\boldsymbol{ \eta } _{k_2}$. The position of each of the ends of these branches is then 
 \[ 
\mathbf{c}_{k_1}=  \boldsymbol{r}(s)+ \Lambda  _1 (s) \boldsymbol{\eta}_{k_1}(s) \, . 
 \]
 Since these second level rigid bouquets can evolve independently we would need $n_1$ variables $\Lambda_2 ^{k_1}(s)$, $k_1=1,...,n_1$ instead of simply $ \Lambda _2(s)$. The position of a charge  at the second level is therefore
 \[
 \mathbf{c}_{k_1 k_2}(s)= 
\boldsymbol{r}(s)+ \Lambda  _1 (s) \boldsymbol{\eta}_{k_1}(s)+ \Lambda _2 ^{k_1} (s)\Lambda _1 (s) \boldsymbol{\eta}_{k_2}(s).
\] 
 Similarly the position of a charge at the $i$-th level depends on the branches at lower level. 
 Thus if we denote by $n_j$ the number of branches at the $j$th level and by $k_j=1,...,n_j$ the indices of these branches, the  variables needed at the $i$th level are $ \Lambda_i ^{k_{i-1},k_{i-2},...,k_1}(s)$ instead of simply $ \Lambda _i (s)$. 
The position of a charge at the $i$-th level is then 
\begin{align*}
\mathbf{c}_{k_1 k_2 \ldots k_n}=\boldsymbol{r}(s)+ & \Lambda  _1 (s) \boldsymbol{\eta}_{k_1}(s)+ \Lambda _2 ^{k_1} \Lambda _1 (s)\boldsymbol{\eta}_{k_2}(s)+ \\ 
& \quad \quad \quad ....+ \Lambda _i^{k_{i-1},....,k_1} (s)\Lambda _{i-1}^{k_{i-2},...,k_1}(s).... \Lambda _2^{k_1}(s) \Lambda _1 (s)  \boldsymbol{\eta}_{k_1}(s)
\end{align*}
 The configuration space is then given by 
\[
SE(3)\,\circledS\,(SO(3)^{n_1}\,\circledS\,(SO(3)^{n_1n_2}\,\circledS\,(\cdots\,\circledS\, SO(3)^{n_1n_2...n_N})\cdots)),
\]
where $SO(3)^n=SO(3)\times ... \times SO(3)$, and the product is taken $n$ times.
Strictly speaking, we should include the variables $\Lambda^{k_{i-1} \ldots k_1}_i$ in the formulas but this would make the resulting expressions appear unnecessarily complex, especially in the case of $N$-bouquets. Thus, we shall only deal with one variable $\Lambda _i (s)$ at the $i$th level and leave the general case to the interested reader. We hope no confusion will arise from having made this simplification for the sake of streamlining the notation.
\rem{%%%%%%%%%%%%%%%%%%%%%%%%%%%%%%%%%%
\todo{VP: Now I am not sure we need the paragraph below. I leave the fate of that short paragraph to you. It feels a bit out of place to me now.\\
F+T: Either we erase it or we put the graph-like case as an idea for a future direction at the end of the paper.} 
These comments provide another reason why the theory we develop here needs substantial modification if the structure of the dendrites is graph-like, \emph{i.e.}, the branches connect at some later level. However, 
in this paper we only consider tree-like dendrimers as being the only physically feasible. 
}
%%%%%%%%%%%%%%%%%%%%%%%%%%%%%%%%%%
\quad $\blacklozenge$
\end{remark} 

\paragraph{Motivation for the affine Euler-Poincar\'e theory.}
Recall that for the molecular strand (corresponding to a one-bouquet; see \S\ref{review_strand}) the main goal was to treat a nonlocal potential depending on the separation distance
\[
d_{km}(s,s'):=|\boldsymbol{c}_k(s)-\boldsymbol{c}_m(s')|,\qquad\text{where}\qquad \boldsymbol{c}_k(s)=\br(s)+\Lambda(s)\boldeta_k(s).
\]
The crucial property in the implementation of the affine Euler-Poincar\'e theory is the invariance of the expression for $\bd_{km}(s,s')$ under the action of $SO(3)$  given by
\[
(\Lambda,\br)\mapsto (h\Lambda,h\br), \quad h \in SO(3).
\]

For the generalization to two-bouquets, the position of the charge is given by
\[
\boldsymbol{c}_{k_1k_2}(s)=\br(s)+\Lambda_1(s)\boldeta_{k_1}(s)+\Lambda_2(s)\Lambda_1(s)\boldeta_{k_2}(s),
\]
and we need to treat a potential depending on the distances
\[
d_{k_1k_2,m_1m_2}(s,s'):=|\boldsymbol{c}_{k_1k_2}(s)-\boldsymbol{c}_{m_1m_2}(s')|.
\]
In this case, we have $SO(3)$-invariance of the distances under the transformation
\[
(\Lambda_1(s),\br(s),\Lambda_2(s))\mapsto (h\Lambda_1(s),h\br(s),h\Lambda_2(s)h^{-1}), \quad h \in SO(3),
\]
which shows that the action by homomorphisms again arises naturally, as in the case of the coupled rigid bodies. It is therefore appropriate to consider the semidirect product $SE(3) \,\circledS\, SO(3)$. On the other hand, for the simpler case of the molecular strand, we already know that a cocycle is needed to understand geometrically the reduction process from the standard Euler-Lagrange equations to the reduced equations of motion. These observations strongly suggest that the geometric setting developed in Section \ref{sec_geometric_setting} is appropriate for the description of multi-bouquets, as we now show.
\medskip

Let $SE(3)$ act on $SO(3)$ by $( \Lambda_1, \br) \cdot \Lambda_2 : = \Lambda_1 \Lambda_2 \Lambda_1^{-1}$, where $\Lambda_1, \Lambda_2 \in SO(3)$. The semidirect product 
$SE(3)\,\circledS\,SO(3)$
associated to this action by homomorphisms of $SO(3)$ has
therefore the multiplication
\[ (\Lambda_1,\br,\Lambda_2)(\bar{\Lambda}_1,\bar{\br},\bar{\Lambda}_2)=\left(\Lambda_1\bar{\Lambda}_1,\br+\Lambda_1\bar{\br},\Lambda_2\Lambda_1\bar\Lambda_2\Lambda_1^{-1}\right)
\]
and inverse
\[
(\Lambda_1,\br,\Lambda_2)^{-1}=(\Lambda_1^{-1},
-\Lambda_1^{-1}\br,\Lambda_1^{-1}\Lambda_2^{-1}\Lambda_1)\,.
\]

\paragraph{Formulas for the semidirect product.} 
The conjugation in $SE(3) \,\circledS\,SO(3)$ is given by
\begin{align*}
&\operatorname{AD}_{(\Lambda_1, \boldsymbol{r}, \Lambda_2)}(\bar{\Lambda}_1, \bar{\boldsymbol{r}}, \bar{\Lambda}_2)
= (\Lambda_1, \boldsymbol{r}, \Lambda_2)
(\bar{\Lambda}_1, \bar{\boldsymbol{r}}, \bar{\Lambda}_2)
(\Lambda_1, \boldsymbol{r}, \Lambda_2)^{-1}\\
&\qquad = \left(\Lambda_1 \bar{ \Lambda}_1\Lambda_1^{-1}, 
\boldsymbol{r}+ \Lambda_1 \bar{\boldsymbol{r}} -
\Lambda_1 \bar{ \Lambda}_1\Lambda_1^{-1} \boldsymbol{r}, 
\Lambda_2 \Lambda_1 \bar{\Lambda}_2 \bar{\Lambda}_1
\Lambda_1^{-1}\Lambda_2^{-1}\Lambda_1 \bar{ \Lambda}_1\Lambda_1^{-1} \right)
.
\end{align*}
Consequently, the adjoint action on of $SE(3) \,\circledS\,SO(3)$ on
$\mathfrak{se}(3) \,\circledS\,\mathfrak{so}(3)$ is expressed as
\begin{align}
\label{Adjoint_sdp}
\operatorname{Ad}_{(\Lambda_1, \boldsymbol{r}, \Lambda_2)}
&(\omega_1, \boldsymbol{\gamma}, \omega_2) \nonumber\\
&= \left(\Lambda_1 \omega_1\Lambda_1^{-1}, \Lambda_1 \boldsymbol{\gamma} - \Lambda_1 \omega_1\Lambda_1^{-1} \boldsymbol{r}, \Lambda_2 \Lambda_1(\omega_1+ \omega_2)
\Lambda_1^{-1} \Lambda_2^{-1} - \Lambda_1 \omega_1\Lambda_1^{-1} \right)
\end{align}
and the Lie algebra bracket is given in vector form by
\begin{align}
\label{adjoint_sdp}
&\operatorname{ad}_{(\boldsymbol{\omega}_1, \boldsymbol{\gamma}, \boldsymbol{\omega}_2)} 
(\bar{\boldsymbol{\omega}}_1, \bar{\boldsymbol{\gamma}}, \bar{\boldsymbol{\omega}}_2)
=\left[(\boldsymbol{\omega}_1, \boldsymbol{\gamma}, \boldsymbol{\omega}_2), 
(\bar{\boldsymbol{\omega}}_1, \bar{\boldsymbol{\gamma}}, \bar{\boldsymbol{\omega}}_2) \right] \nonumber\\
&\qquad = \left(\boldsymbol{\omega}_1 \times 
\bar{\boldsymbol{\omega}}_1, 
\boldsymbol{\omega}_1 \times 
\bar{\boldsymbol{\gamma}}  
- \bar{\boldsymbol{\omega}}_1 \times 
\boldsymbol{\gamma},
\boldsymbol{\omega}_2 \times 
\bar{\boldsymbol{\omega}}_2
+ \boldsymbol{\omega}_1 \times 
\bar{\boldsymbol{\omega}}_2
+ \boldsymbol{\omega}_2 \times 
\bar{\boldsymbol{\omega}}_1
\right).
\end{align}
Therefore, the infinitesimal coadjoint action then reads
\begin{align}
\label{adstar_sdp}
\!\!\!
\operatorname{ad}_{(\boldsymbol{\omega}_1, \boldsymbol{\gamma}, \boldsymbol{\omega}_2)}^\ast
(\boldsymbol{\mu}_1, \boldsymbol{\nu}, \boldsymbol{\mu}_2)
= \left(\boldsymbol{\mu}_1 \times\boldsymbol{\omega}_1 - 
\boldsymbol{\gamma} \times \boldsymbol{\nu} + 
\boldsymbol{\mu}_2 \times \boldsymbol{\omega}_2, 
\boldsymbol{\nu} \times \boldsymbol{\omega}_1,
\boldsymbol{\mu}_2 \times (\boldsymbol{\omega}_1
+\boldsymbol{\omega}_2)
\right).
\end{align}

\paragraph{The reduced variables.} For the problem of two-level bouquets
the configuration space $\mathcal{F}(I,SE(3)\,\circledS\,
SO(3))$
consists of smooth functions of an interval $I $ with values in $SE(3)\,\circledS\,SO(3)$.
This is a group relative to pointwise multiplication and
hence we may apply the formulas derived above. To simplify
notation, we use the same symbols $(\Lambda_1, \boldsymbol{r}, \Lambda_2)$ to denote variables in $\mathcal{F}(I,SE(3)\,\circledS\,SO(3))$. These are functions depending on 
the curve parameter $s \in I$.

The theory developed in Section \ref{red_equ_motion} now applies. Take $G_1 = \mathcal{F}(I, SE(3)) \ni ( \Lambda_1, \boldsymbol{r})$,
$G_2 = \mathcal{F}(I, SO(3)) \ni \Lambda_2$, and the action of $G_1$ on $G_2 $ by group homomorphisms given by $(\Lambda_1, \br) \cdot \Lambda_2 : = \Lambda_1 \Lambda_2 \Lambda_1^{-1}$. The representation space is 
$ V= V ^\ast =\mathcal{F}(I, \mathbb{R}^3)^4\ni (
\boldsymbol{\Omega}_1, \boldsymbol{\Gamma}, \boldsymbol{\Omega}_2, \boldsymbol{\rho})$ where we identify $V$ and
$V ^\ast$ by the $L^2$-inner product on $I$. The
$\mathcal{F}(I, SE(3)) \,\circledS\,\mathcal{F}(I, SO(3))$-representation on $\mathcal{F}(I, \mathbb{R}^3)^4$ is
given in vector form by 
\begin{align*}
(\Lambda_1, \br, \Lambda_2)(\boldsymbol{\Omega}_1, 
\boldsymbol{\Gamma}, \boldsymbol{\Omega}_2, 
\boldsymbol{\rho}) : 
&= \left(\operatorname{Ad}_{(\Lambda_1, \br, \Lambda_2)}
(\boldsymbol{\Omega}_1, \boldsymbol{\Gamma}, 
\boldsymbol{\Omega}_2), \Lambda_1\boldsymbol{\rho} \right)\\
& = \left(\Lambda_1\boldsymbol{\Omega}_1, \Lambda_1 
\boldsymbol{\Gamma} - \Lambda_1 \boldsymbol{\Omega}_1
\times \boldsymbol{r}, \Lambda_2\Lambda_1
(\boldsymbol{\Omega}_1 + \boldsymbol{\Omega}_2) - 
\Lambda_1 \boldsymbol{\Omega}_1, \Lambda_1\boldsymbol{\rho}
\right)
\end{align*}
and the group one-cocycle $c$ on $\mathcal{F}(I, SE(3)) \,\circledS\,\mathcal{F}(I, SO(3))$ is taken as 
\begin{equation}
\label{concrete_cocycle}
c(\Lambda_1,\br,\Lambda_2):=\left((\Lambda_1,\br,\Lambda_2)\partial_s(\Lambda_1,\br,\Lambda_2)^{-1},-\br\right).
\end{equation}
The choice of this cocycle is motivated by a convenient
expression of the reduced variables, as we shall see below.

Indeed, using the formula of the quotient map $T\left(\mathcal{F}(I, SE(3)) \,\circledS\,\mathcal{F}(I, SO(3)) \right) \times \mathcal{F}(I, \mathbb{R}^3) ^4 \rightarrow \mathcal{F}(I, \mathfrak{se}(3)) \times \mathcal{F}(I, SO(3)) \times \mathcal{F}(I, \mathfrak{so}(3)) \times \mathcal{F}(I, \mathbb{R}^3) ^4$ given in \eqref{2_bouquet_quotient_map} with $a_0=0$, we obtain
\begin{align*}
(\Lambda_1,\dot\Lambda_1,\br,\dot\br,\Lambda_2,\dot\Lambda_2)&\mapsto \left((\Lambda_1,\br,\Lambda_2)^{-1}(\Lambda_1,\dot\Lambda_1,\br,\dot\br,\Lambda_2,\dot\Lambda_2),\Lambda_1^{-1}\Lambda_2\Lambda_1,c((\Lambda_1,\br,\Lambda_2)^{-1})\right)\\
&=(\omega_1,\bgam,\omega_2,p,\Omega_1,\bGam,\Omega_2,\brho),
\end{align*}
where
\begin{align*}
(\omega_1,\bgam,\omega_2):&=(\Lambda_1,\br,\Lambda_2)^{-1}(\Lambda_1,\dot{\Lambda}_1,\br,\dot{\br},\Lambda_2,\dot{\Lambda}_2)\\
&=(\Lambda^{-1},-\Lambda^{-1}\br,\Lambda_1^{-1}\Lambda_2^{-1}\Lambda_1)(\Lambda_1,\dot{\Lambda}_1,\br,\dot{\br},\Lambda_2,\dot{\Lambda}_2)\\
&=\left(\Lambda_1^{-1}\dot{\Lambda}_1,\Lambda_1^{-1}\dot{\br},\Lambda_1^{-1}\Lambda_2^{-1}\dot{\Lambda}_2\Lambda_1\right),
\end{align*}
and
\begin{align*}
(\Omega_1,\bGam,\Omega_2,\brho):&=c\left((\Lambda_1,\br,\Lambda_2)^{-1}\right)=\left((\Lambda_1,\br,\Lambda_2)^{-1}\partial_s(\Lambda_1,\br,\Lambda_2),\Lambda_1^{-1}\br\right)\\
&=\left(\Lambda_1^{-1}\Lambda'_1,\Lambda_1^{-1}\br',\Lambda_1^{-1}\Lambda_2^{-1}\Lambda'_2\Lambda_1,\Lambda_1^{-1}\br\right).
\end{align*}
We continue to use the convention that for a vector $\mathbf{v} \in \mathbb{R}^3$ we denote by $v : = \widehat{\mathbf{v}}\in \mathfrak{so}(3)$ the matrix defined by $v \mathbf{w} =
\mathbf{v} \times \mathbf{w}$.
Summarizing, we have defined the reduced convective variables
\begin{equation}
\label{reduced_variables_list}
\begin{array}{llllll}
\omega_1&=&\Lambda_1^{-1}\dot\Lambda_1
&
\Omega_1&=&\Lambda_1^{-1}\Lambda'_1\\
\bgam&=&\Lambda_1^{-1}\dot\br
&
\bGam&=&\Lambda_1^{-1}\br'\\
\omega_2&=&\Lambda_1^{-1}\Lambda_2^{-1}\dot{\Lambda}_2\Lambda_1 \qquad \qquad
&
\Omega_2&=&\Lambda_1^{-1}\Lambda_2^{-1}\Lambda'_2\Lambda_1\\
p&=&\Lambda_1^{-1}\Lambda_2\Lambda_1
&
\brho&=&\Lambda_1^{-1}\br.
\end{array}
\end{equation}
These variables will be shown to be well adapted to the
description of the dynamics of two-bouquets. Here, $\omega_i$ are angular rotations of each bouquet, with $\omega_1$
being measured with respect to the base coordinate system and $\omega_2$ measured with respect to $\Lambda_1$. 
Similarly, $\Omega_1$ and $\bGam$ are the familiar variables describing the infinitesimal twist and stretching of the base, whereas 
$\Omega_2$ describes the infinitesimal twist of the second bouquet in the coordinate frame connected with the first 
bouquet. Notice that there is no $\bGam_2$ as there is no stretching associated with the second bouquet. 
Finally, the new variable $p$ is the orientation of the second bouquet seen from the coordinate frame of the first bouquet. 
Thus, $p$ is a new variable associated with the multi-bouquet structure of the polymer.

\paragraph{The Lagrangian for two-bouquets.}
In this case, the material Lagrangian depends only on
the variables $(\Lambda_1,\dot{\Lambda}_1, \boldsymbol{r}, \dot{ \boldsymbol{r}},\Lambda_2,\dot{\Lambda}_2)$. It is 
important to interpret it as being the 
restriction of a parameter-dependent Lagrangian defined
on $T \mathcal{F}(I, SE(3) \,\circledS\,SO(3)) \times 
\mathcal{F}(I, \mathbb{R}^3)^4$ for the zero value of the
parameter in $\mathcal{F}(I, \mathbb{R}^3)^4$, that is,
\[
L_{(a_0=0)} = L(\Lambda_1,\dot{\Lambda}_1, \boldsymbol{r}, \dot{ \boldsymbol{r}},\Lambda_2,\dot{\Lambda}_2).
\]
We are thus in the situation described by the theory developed in \S\ref{sec_fixed_par}, provided $L $
is $(G _1)^c_0$-invariant. We have $(G _1)^c_0= SO(3)
\subset \mathcal{F}(I, SE(3))$ as a direct verification
using \eqref{concrete_cocycle} shows.

\paragraph{Nonlocal variables.} 
The most general form of the symmetry-reduced Lagrangian for level-two bouquets is 
\[
\ell=\ell_{loc}(\bom_1, \bgam, \bom_2, p,\bOm_1,\bGam,\bOm_2,\brho) + 
\ell_{np}(\bom_1, \bgam, \bom_2, p,\bOm_1,\bGam,\bOm_2,\brho, (\Lambda_1, \br)),
\]
where the first Lagrangian $\ell_{loc}$ is explicitly given in terms of the variables $\bom_1$, $\bgam$, $\bom_2$, $p$,
$\bOm_1$, $\bGam$, $\bOm_2$, $\brho$ and the second Lagrangian $\ell_{np}$ has still a dependence on $(\Lambda_1, \br)$, where $(\Lambda_1, \br)$ are such that $c((\Lambda_1, \br)^{-1}, p ^{-1}) = 
(\bOm_1,\bGam,\bOm_2,\brho)$, that is, the definition 
\eqref{reduced_variables_list} of the
reduced variables holds. More precisely, $\ell_{np}$ has
the nonlocal expression
\begin{equation} \label{l_np}
\ell_{np} = \iint U \big(\xi(s,s'), \bkappa(s,s'), p(s),
p(s'), \bGam(s), \bGam(s')\big) \mbox{d} s \mbox{d} s', 
\end{equation} 
where $U: SO(3) \times (\mathbb{R}^3)^5 \rightarrow 
\mathbb{R}$ is a given function and
\begin{align*} 
\bkappa(s,s')&:=-\Lambda_1^{-1} (s) \big(\br(s)-\br(s') \big)
\in \mathbb{R}^3,\\ 
\xi(s,s')&:=\Lambda_1^{-1} (s) \Lambda_1(s') \in \mathfrak{so}(3). 
\end{align*}
This general expression of $\ell_{np}$ allows the treatment
of a potential depending on the Euclidean distance between 
two charges. Indeed, as we can see in Figure 
\ref{fig:multi-bouquet},
the position of a charge is given by
\[
\boldsymbol{c}_{k_1k_2}(s)=\br(s)+\Lambda_1(s)\boldeta_{k_1}(s)+\Lambda_2(s)\Lambda_1(s)\boldeta_{k_2}(s),
\]
and so the distance between two charges is found to be
\begin{align}
\label{d12}
&d_{k_1 k_2 m_1 m_2}(s,s') =
\left| 
\boldsymbol{c}_{k_1k_2}(s) - \boldsymbol{c}_{k_1k_2}(s') 
\right| 
\nonumber 
\\
&=
\left| 
\br(s)+\Lambda_1(s)\boldeta_{k_1}(s)+\Lambda_2(s)\Lambda_1(s)\boldeta_{k_2}(s)
- \br(s')-\Lambda_1(s')\boldeta_{m_1}(s')-\Lambda_2(s')\Lambda_1(s')\boldeta_{m_2}(s')
\right|
\nonumber 
\\ 
&= 
\left| 
\Lambda_1^{-1} (s) \big( \br(s)-\br(s') \big)+\boldeta_{k_1}(s)+\Lambda_1^{-1} (s) \Lambda_2(s)\Lambda_1(s)\boldeta_{k_2}(s)
-
\right.
\nonumber 
\\ 
&\quad \quad \quad 
\left.
\Lambda_1^{-1} (s) \Lambda_1(s')\boldeta_{m_1}(s')
-\Lambda_1^{-1} (s) \Lambda_2(s')\Lambda_1(s')\boldeta_{m_2}(s')
\right| 
\nonumber 
\\ 
&=
\left| 
-\bkappa(s,s') + \boldeta_{k_1}(s)+p(s) \boldeta_{k_2}(s) 
-
\xi(s,s') \boldeta_{m_1}(s') 
-\xi(s,s') p(s') \boldeta_{m_2}(s') 
\right|. 
\end{align}
This expression shows that the distance between two 
charges can be expressed solely in terms of the variables $\xi(s, s'), 
\boldsymbol{\kappa}(s, s'), p (s),  p (s')$. In addition,
we note that $\ell_{np}$ is invariant under the 
$(G _1)^c_0= SO(3)$-action
\[
(\Lambda_1(s),\br(s),\Lambda_2(s))\mapsto (h\Lambda_1(s),h\br(s),h\Lambda_2(s)h^{-1}), \quad h \in SO(3).
\]
Hence, we are in the situation described in \S
\ref{sec_fixed_par}. 

\paragraph{Reduced equations of motion.} In order to write the reduced equations \eqref{reduced_EL}, we need to consider the affine representation $\theta$ of the group $G=\mathcal{F}(I,SE(3)\,\circledS\,SO(3))$ on $V^*=\Omega^1(I,\mathfrak{se}(3)\,\circledS\,\mathfrak{so}(3))\times\mathcal{F}(I,\mathbb{R}^3)$, given by
\[
\theta_{(\Lambda_1,\boldsymbol{r},\Lambda_2)}(\boldsymbol{\Omega}_1,\boldsymbol{\Gamma},\boldsymbol{\Omega}_2,\boldsymbol{\rho})=(\Lambda_1,\boldsymbol{r},\Lambda_2)\left(\boldsymbol{\Omega}_1,\boldsymbol{\Gamma},\boldsymbol{\Omega}_2,\boldsymbol{\rho}\right)+c(\Lambda_1,\boldsymbol{r},\Lambda_2),
\]
where the first term denotes the representation of $G$ on $V^*$ defined by
\begin{align*}
(\Lambda_1,r,\Lambda_2)(\boldsymbol{\Omega}_1,\boldsymbol{\Gamma},\boldsymbol{\Omega}_2,\boldsymbol{\rho})
&= \left(\Lambda_1\boldsymbol{\Omega}_1, \Lambda_1 
\boldsymbol{\Gamma} - \Lambda_1 \boldsymbol{\Omega}_1
\times \boldsymbol{r}, \Lambda_2\Lambda_1
(\boldsymbol{\Omega}_1 + \boldsymbol{\Omega}_2) - 
\Lambda_1 \boldsymbol{\Omega}_1, \Lambda_1\boldsymbol{\rho}
\right).
\end{align*}
Thus, upon using the abstract system \eqref{reduced_EL}, the
reduced Euler-Lagrange equations become
\begin{equation}
\label{dynamic_gen} 
\!\!\!\!\!\!\!\!
\left\{
\begin{aligned}
&\Big(\partial_t + \bom_1\times \Big)\dede{\ell}{\bom_1} 
+ \Big(\partial_s + \bOm_1\times\Big)\dede{\ell}{\bOm_1} 
+\brho\times\dede{\ell}{\brho}
+\bGam\times\dede{\ell}{\bGam}
+\bgam\times\dede{\ell}{\bgam}  \\
&\qquad \qquad  \qquad 
+\bom_2\times\dede{\ell}{\bom_2}
+\bOm_2\times\dede{\ell}{\bOm_2}
+ \frac{\delta \ell}{\delta p} p^{-1}
- p^{-1}\frac{\delta \ell}{\delta p}  =0 \\
&\Big( \prt_t + \bom_1 \times\Big)
 \dede{\ell}{\bgam} + 
 \Big(\prt_s + \bOm_1\times\Big)
 \dede{\ell}{\bGam} -\dede{\ell}{\brho}  =0 \\
&\Big( \prt_t + \bom_2\times\Big)\dede{\ell}{\bom_2} 
+ \Big(\prt_s + \bOm_2\times\Big)\dede{\ell}{\bOm_2}+
\bom_1\times\frac{\delta \ell}{\delta\bom_2}+
\bOm_1\times\frac{\delta \ell}{\delta\bOm_2}
- p^{-1}\frac{\delta \ell}{\delta p}  =0 .
\end{aligned}
\right.
\end{equation}
In these equations the functional partial derivatives
are total derivatives. These equations could have been
obtained also directly by using the explicit expressions
of the variations for all variables, namely,
\begin{align*}
&\delta\omega_1=\frac{\partial}{ \partial t} \Sigma_1 + [\omega_1, \Sigma_1], \qquad \; \qquad
\delta\bgam=\frac{\partial}{ \partial t} \boldsymbol{\Psi}
 + \boldsymbol{\omega}_1\times \boldsymbol{\Psi} + \boldsymbol{\gamma}\times  \boldsymbol{\Sigma}_1\\
&\delta\omega_2= \frac{\partial}{ \partial t} \Sigma_2 + 
[ \Sigma_1, \omega_2] + [\Sigma_2, \omega_1] +
[ \Sigma_2, \omega_2], \qquad \; \qquad
\delta p=p \Sigma_1 - \Sigma_1 p + p \Sigma_2\\
& \delta\Omega_1=\frac{\partial}{ \partial s} \Sigma_1 + [\Omega_1, \Sigma_1], \qquad\qquad
\delta\bGam=\frac{\partial}{ \partial s} \boldsymbol{\Psi}
 + \boldsymbol{\Omega}_1\times \boldsymbol{\Psi} + \boldsymbol{\Gamma}\times  \boldsymbol{\Sigma}_1\\
 &
\delta\Omega_2=\frac{\partial}{ \partial s} \Sigma_2 + 
[ \Sigma_1, \Omega_2] + [\Sigma_2, \Omega_1] +
[ \Sigma_2, \Omega_2], \qquad \qquad
\delta\brho=\boldsymbol{\rho} \times \boldsymbol{\Sigma}_1 + \boldsymbol{\Psi}.
\end{align*}
where $\Sigma_1 := \Lambda_1^{-1}\delta\Lambda_1$, $\boldsymbol{\Psi}:= \Lambda_1^{-1}\delta\boldsymbol{r}$,
$\Sigma_2 := \Lambda_1^{-1} \Lambda_2^{-1} \delta\Lambda_2 \Lambda_1$.

If we use the explicit
expression of $\ell=\ell_{loc}+\ell_{np}$, with $\ell_{np}$ given in \eqref{l_np}, this system becomes
\begin{equation}
\label{dynamic}
\left\{
\begin{aligned}
&\Big(\prt_t + \bom_1\times \Big) \dede{\ell_{loc}}{\bom_1} 
+ \Big(\prt_s + \bOm_1\times\Big)\dede{\ell_{loc}}{\bOm_1} 
+ \brho\times\dede{\ell_{loc}}{\brho}
+ \bGam\times\dede{(\ell_{loc} + \ell_{np})}{\bGam} \\
&\qquad \qquad 
+ \bgam\times\dede{\ell_{loc}}{\bgam}
+ \bom_2\times\dede{\ell_{loc}}{\bom_2}
+\bOm_2\times\dede{\ell_{loc}}{\bOm_2}
+ \frac{\delta \ell_{loc}}{\delta p} p^{-1}
- p^{-1}\frac{\delta \ell_{loc}}{\delta p}\\
& \qquad \qquad +\int\left(\frac{\partial U}{\partial p_1}(s,s')p(s)^{-1} - p(s)^{-1}\frac{\partial U}{\partial p_1}(s,s') 
\right) \mbox{d} s' \\
&\qquad \qquad
+\int\left(\frac{\partial U}{\partial p_2}(s',s)p(s)^{-1} - p(s)^{-1}\frac{\partial U}{\partial p_2}(s',s)
\right) \mbox{d} s' \\
& \qquad \qquad -\int \left( \frac{\partial U}{\partial \bkappa} (s,s') \times \bkappa (s,s')+ \mathbf{Z}(s,s') \right) 
 \mbox{d} s' =0,
  \\
&\Big( \prt_t + \bom_1  \times\Big)
 \dede{\ell_{loc}}{\bgam} + 
 \Big(\prt_s + \bOm_1\times\Big)
 \dede{(\ell_{loc}+\ell_{np})}{\bGam}
 -\dede{\ell_{loc}}{\brho} 
 \\
& \qquad \qquad  -\int \left(
\xi(s,s') \frac{\partial U}{\partial \bkappa} (s',s)
-\frac{\partial U}{\partial \bkappa} (s,s')
\right) \mbox{d} s' =0,
 \\
&\Big( \prt_t + \bom_2\times\Big)\dede{\ell_{loc}}{\bom_2} 
+ \Big(\prt_s + \bOm_2\times\Big)\dede{\ell_{loc}}{\bOm_2}+
\bom_1\times\frac{\delta \ell_{loc}}{\delta\bom_2}+
\bOm_1\times\frac{\delta \ell_{loc}}{\delta\bOm_2} \\
&\qquad \qquad  
- \int \left( p(s)^{-1}\frac{\partial U}{\partial p_1}(s,s') 
+ p(s)^{-1}\frac{\partial U}{\partial p_2}(s',s)\right) 
\mbox{d} s' 
=0, 
\end{aligned}
\right.
\end{equation}
where the functional derivatives of $\ell_{loc}$ and $\ell_{np}$ are now usual ones and
the term $\bZ(s,s') \in \mathbb{R}^3$ is the vector corresponding to
\begin{equation*}
\widehat\bZ(s,s') = \xi(s,s')
\frac{\partial U}{\partial \xi} (s',s)
-\frac{\partial U}{\partial \xi} (s,s')
\xi^T(s,s') \in \mathfrak{so}(3).
\end{equation*}
The expressions $\partial U/\partial \xi$, $\partial U/\partial\boldsymbol{\kappa}$, $\partial U/\partial p_1$, and $\partial U/\partial p_i$ denote the partial derivative of  $U=U(\xi,\boldsymbol{ \kappa} , p_1, p_2)$ as a function on $SO(3)\times \mathbb{R} ^3 \times SO(3)\times SO(3)$. We identify the cotangent space $T_A^*SO(3)$ with $T_ASO(3)$ by using the pairing
\begin{equation} \label{pairing_SO3}
\left\langle V_A,U_A \right\rangle :=\mathbf{v}\cdot\mathbf{u}, \quad\text{for}\quad  \widehat{\mathbf{v}}=A^{-1}V_A,\;\;\widehat{\mathbf{u}}=\cdot A^{-1}U_A.
\end{equation}
Using this identification, we have $\partial U/\partial\xi\in T_\xi SO(3)$ and $\partial /\partial p_i\in T_{p_i}SO(3)$.

To these equations one needs to add  the kinematic equations
\eqref{kinematic_equations_abstract} which in this case take the forms
\[
\left\{\begin{array}{l}
\vspace{0.2cm}\displaystyle\dot p=p\bom_2-\bom_1 p+p\bom_1\\
\vspace{0.2cm}\displaystyle(\partial_t+\bom_1\times)\bOm_1=\partial_s\bom_1\\
\vspace{0.2cm}\displaystyle(\partial_t+\bom_1\times)\bGam=(\partial_s+\bOm_1\times)\bgam\\
\vspace{0.2cm}\displaystyle(\partial_t+(\bom_1+\bom_2)\times)\bOm_2=(\partial_s+\bOm_1\times)\bom_2\\
\vspace{0.2cm}\displaystyle(\partial_t+\bom_1\times)\brho=\bgam.
\end{array}
\right.
\]
Note that the terms involving $\xi$ and $\bkappa$ do not enter the angular momentum for the second bouquet. This 
is because there is no term analogous to $\xi$ involving 
$\Lambda_2$. 

The Poisson brackets for the Hamiltonian evolution of two-bouquets may be obtained from the general formula given in \S\ref{Hamiltonian_formulation}.

\subsection{Computation of derivatives for the nonlocal term} 

While equations (\ref{dynamic}) define the complete equations of motion for a two-level bouquet string, for practical applications it is useful to provide explicit formulas for the partial derivatives in that equation and clarify their physical meaning. The terms 
\[ 
\frac{\de \ell_{loc} }{\de \bom_1} \, , \quad \frac{\de \ell_{loc} }{\de \bgam} \, ,
\] 
are the angular and linear momenta in the coordinate frame associated with the filament. These momenta are connected to  $\bom_1$ and $\bgam$ by the geometry of the filament and are usually linear functions of $\bom_1$ and $\bgam$. Similarly, 
\[ 
\frac{\de \ell_{loc} }{\de \bOm_1} \, , \quad \frac{\de \ell_{loc} }{\de \bGam_1}\,  ,
\] 
are the elastic deformations associated with twisting and stretching of the filament. These quantities are usually related 
 by the constitutive relations to the deformations themselves, which defines the physics of the base elastic rod. The terms involving the derivatives 
 \[ 
 \frac{\de \ell_{loc} }{\de \bom_2} \, , \quad \frac{\de \ell_{loc} }{\de \bOm_2} \, ,
 \] 
 are the angular momentum and torque applied to the bouquet, respectively, and are related by physical and geometric considerations to 
 ($\bom_i$, $\bgam$) for the first term and to ($\bOm_i$, $\bGam$) for the second term.   In any case, the derivatives of the 
 local terms should not cause any mathematical difficulty (although they may require quite detailed physical considerations). 
 
 More care must be exercised in computing the derivatives with respect to the nonlocal variables $(\xi, \bkappa, p)$. The derivatives of the local (elastic) Lagrangian 
 \[ 
  \frac{\de \ell_{loc} }{\de p} p^{-1} \, , \quad p  \frac{\de \ell_{loc} }{\de p} 
 \] 
 denote the elastic deformations of second level of bouquets with respect to the first level. The $p$-dependence of the local Lagrangian may also come from geometric factors affecting moment of inertia of a two-bouquet, or total potential energy 
 of the deformation. Without considering a particular molecule, these terms are difficult to identify. On the other hand, using the procedure outline below, these derivatives should be relatively easy to compute once the functional dependence of $\ell_{loc}$ on $p$ is known.  

 We therefore concentrate on the derivatives in the nonlocal term, which is both very instructive and can be applied to a general potential in view of concrete applications.
\medskip

 We are now ready to compute the variations in the nonlocal terms. 
 For convenience, we define the vector  
 \begin{align} 
 &\mathbf{d}_{k_1 k_2 m_1 m_2} (s,s') \nonumber 
  \\ 
 &\quad =
-\bkappa(s,s') + \boldeta_{k_1}(s)+p(s) \boldeta_{k_2}(s) 
-
\xi(s,s') \boldeta_{m_1}(s') 
-\xi(s,s') p(s') \boldeta_{m_2}(s')  
 \end{align} 
 with 
 \begin{equation} 
d_{k_1 k_2 m_1 m_2} (s,s') = \left|  \mathbf{d}_{k_1 k_2 m_1 m_2} (s,s') \right| \, . 
 \label{ddef2}
 \end{equation} 
 In applications, the potential energy is a function of the distances $d_{k_1 k_2 m_1 m_2} (s,s')$ between any two given charges. More precisely, if $U(d)$ is such a potential (that could be electrostatic, screened electrostatic, Lennard-Jones, or any 
 combinations of those or any other potentials), then the nonlocal part of the Lagrangian is given by 
 \begin{equation} 
 \ell_{np} = \sum_{k_1, k_2, m_1, m_2} \int U\big( d_{k_1 k_2 m_1 m_2} (s,s') \big) 
 \left|  \bGam(s) \right|  \left|  \bGam(s') \right| \mbox{d} s \,\mbox{d} s' \, . 
 \label{lnp} 
 \end{equation} 
The reverse distance between the points $s$ and $s'$ is
determined by the following formula:
 \begin{align} \label{dss'_to_ds's}
 &\mathbf{d}_{k_1 k_2 m_1 m_2} (s',s) \nonumber 
  \\ 
 &=
\xi^{-1} (s,s') \bkappa(s,s') + \boldeta_{k_1}(s')+p(s') \boldeta_{k_2}(s') 
-
\xi^{-1} (s,s') \boldeta_{m_1}(s) 
-\xi^{-1} (s,s') p(s) \boldeta_{m_2}(s)  
\nonumber
\\ 
&=- \xi^{-1}(s,s')  \mathbf{d}_{m_1 m_2 k_1 k_2} (s,s') 
 \end{align} 
The derivatives with respect to $\xi$ and $p$ are computed using the following lemma.

\begin{lemma} Let $f:SO(3)\rightarrow\mathbb{R}$ be the function defined by
\[
f(A)=\frac{1}{2}|\mathbf{b}-A\mathbf{a}|^2,
\]
where $\mathbf{a}, \mathbf{b}\in \mathbb{R}  ^3 $ are constant vectors. Then the partial derivative of $f$, relative to the pairing \eqref{pairing_SO3}, is given by
\[
\frac{\partial f}{\partial A}=( \mathbf{b} \times A  \mathbf{a} \big) \, \,\widehat{}  \,\, A=( \mathbf{d} \times A  \mathbf{a} \big) \, \,\widehat{}\,\, A,
\]
where $\mathbf{d}=\mathbf{b}-A\mathbf{a}$.
\end{lemma}
 \textbf{Proof.}  Given $V_A\in T_ASO(3)$ and $\widehat{\mathbf{v}}=A ^{-1} V _A $, let $C(\varepsilon)$ be a smooth curve satisfying $C(0) = A $ and $C'(0) = V_A$. We have
 \begin{align*} 
 \left\langle \frac{\partial  f}{\partial A}, V_A \right\rangle
&= \left.\frac{d}{dt}\right|_{ \varepsilon =0}f(C( \varepsilon ))=-\left(\mathbf{b}- A\mathbf{a}\right)\cdot V_A\mathbf{a}=-\left(A ^{-1}\mathbf{b}- \mathbf{a}\right) \cdot 
(\mathbf{v}\times   \mathbf{a})\\
&= (A^{-1}\mathbf{b}\times\mathbf{a})\cdot\mathbf{v}.
\end{align*} 
Therefore, from the definition of the duality pairing \eqref{pairing_SO3}, we conclude that
\[
\frac{\partial f}{\partial A}=A(A^{-1}\mathbf{b}\times\mathbf{a})\,\,\widehat{}\,\,=(\mathbf{b}\times A\mathbf{a})\,\,\widehat{}\,\,A,
\]
 as desired.$\qquad \blacksquare$
 
 \medskip
 
 Using this formula with $\mathbf{d}=\mathbf{d}_{k_1 k_2 m_1 m_2} (s,s')$ and $A=\xi(s,s')$, we deduce that the partial derivative of $U$ relative to $\xi $ is given by
 \begin{align} 
& \frac{\partial U}{\partial \xi} (s,s') 
\nonumber 
\\  
 &= 
 \sum_{ k_1 k_2 m_1 m_2} \frac{U'\big(d_{k_1 k_2 m_1 m_2}(s,s') \big)}{d_{k_1 k_2 m_1 m_2}(s,s')} 
  \Big( 
\mathbf{d}_{k_1 k_2 m_1 m_2} \times 
 \xi\, \big(\boldeta_{m_1}(s') +p(s') \boldeta_{m_2}(s') \big) 
 \Big) 
 \, \, \widehat{} \, \, \xi,
 \label{dUdxi} 
 \end{align}
 where $\mathbf{d}_{k_1 k_2 m_1 m_2}$ and $\xi$ are evaluated at $(s,s')$. Using \eqref{dss'_to_ds's}, we get
 \begin{align} 
 & \frac{\partial U}{\partial \xi} (s',s)
 \nonumber 
 \\ 
 &= 
 \sum_{ k_1 k_2 m_1 m_2} \frac{U'\big(d_{k_1 k_2 m_1 m_2}(s',s) \big)}{d_{k_1 k_2 m_1 m_2}(s',s)} 
  \Big( 
- \xi^{-1}\mathbf{d}_{m_1 m_2 k_1 k_2} \times 
 \xi^{-1}\big(\boldeta_{m_1}(s) +p(s) \boldeta_{m_2}(s) \big) 
 \Big) 
 \, \, \widehat{} \, \, \xi^{-1}
 \nonumber 
 \\ 
 &= 
 \sum_{ k_1 k_2 m_1 m_2} \frac{U'\big(d_{k_1 k_2 m_1 m_2}(s',s) \big)}{d_{k_1 k_2 m_1 m_2}(s',s)} 
 \xi^{-1} \Big( 
-  \mathbf{d}_{m_1 m_2 k_1 k_2}\times 
    \big(\boldeta_{m_1}(s) +p(s) \boldeta_{m_2}(s) \big) 
 \Big) 
 \, \, \widehat{} \,\,,
 \label{dUdxi2} 
 \end{align} 
where $\mathbf{d}_{m_1 m_2 k_1 k_2}$ and $\xi^{-1}$ are evaluated at $(s,s')$.  Thus the vector $\bZ(s,s') $ appearing in the equations of motion can be explicitly computed using the formulas
\begin{align*} 
 & \frac{\partial U}{\partial \xi}(s,s') \xi^{-1}(s,s') \\
 &\qquad \qquad= 
 \sum_{ k_1 k_2 m_1 m_2} \frac{U'\big(d_{k_1 k_2 m_1 m_2}(s,s') \big)}{d_{k_1 k_2 m_1 m_2}(s,s')} 
  \Big( 
\mathbf{d}_{k_1 k_2 m_1 m_2} \times 
 \xi\,  \big(\boldeta_{m_1}(s') +p(s') \boldeta_{m_2}(s') \big) 
 \Big) 
 \, \, \widehat{}
 \nonumber 
 \\ 
 & \xi(s,s') \frac{\partial U}{\partial \xi}(s',s)\\
 &\qquad \qquad=  
 \sum_{ k_1 k_2 m_1 m_2} \frac{U'\big(d_{k_1 k_2 m_1 m_2}(s',s) \big)}{d_{k_1 k_2 m_1 m_2}(s',s)} 
 \Big( 
-  \mathbf{d}_{m_1 m_2 k_1 k_2} \times 
    \big(\boldeta_{m_1}(s) +p(s) \boldeta_{m_2}(s) \big) 
 \Big) 
 \, \, \widehat{} \,\,.
 \end{align*}
 Terms involving derivatives with respect to $p$ are computed similarly. We are not going to present them all. For example, 
 \begin{align*} 
  &p^{-1}(s)  \frac{\partial U}{\partial p(s)}(s,s') = 
 \sum_{ k_1 k_2 m_1 m_2} \frac{U'\big(d_{k_1 k_2 m_1 m_2}(s,s') \big)}{d_{k_1 k_2 m_1 m_2}(s,s')} 
\Big( p^{-1}(s)  \mathbf{d}_{k_1 k_2 m_1 m_2}(s,s') \times \boldeta_{k_2}(s) \Big) \, \, \widehat{} 
\nonumber 
\\ 
&p^{-1}(s)  \frac{\partial U}{\partial p(s')}(s',s) \\
 &\qquad \qquad= 
 \sum_{ k_1 k_2 m_1 m_2} \frac{U'\big(d_{k_1 k_2 m_1 m_2}(s,s') \big)}{d_{k_1 k_2 m_1 m_2}(s,s')} 
\Big(- p^{-1}(s) \mathbf{d}_{m_1 m_2 k_1 k_2}(s,s') \times \boldeta_{m_2}(s) \Big) \, \, \widehat{} \,\,\,.
 \end{align*}

\section{Generalization to $N$-bouquets}

We shall now turn our attention to the general case of $N$-bouquets. For the generalization to $N$-bouquets, the position of the charge is given by
\[
\boldsymbol{c}_{k_1k_2...k_N}(s)=\br(s)+\Lambda_1(s)\boldeta_{k_1}(s)+\Lambda_2(s)\Lambda_1(s)\boldeta_{k_2}(s)+...+\Lambda_N(s)...\Lambda_2(s)\Lambda_1(s)\boldeta_{k_N}(s),
\]
where $\boldsymbol{r}(s)$ is the parametrization of the filament, $\boldsymbol{\eta}_{k_i}(s)$ is a vector of constant length that determines the position of the
the branches at level $i$, where $k_i$ are the labels
of the branches at this level, all of this in the reference
configuration. The proper rotation matrices $\Lambda_i(s) $
give the orientation of the $i$th branch at parameter
value $s$. Generalizing the case of $2$-bouquets, the potential
depends on the distances
\[
\bd_{k_1k_2..k_N,m_1m_2...m_N}(s,s'):=|\boldsymbol{c}_{k_1k_2...k_N}(s)-\boldsymbol{c}_{m_1m_2...m_N}(s')|
\]
between the charges $(k_1,k_2,\ldots, k_N)$ and $(m_1,m_2,\ldots, m_N)$ on the filament at position $s$ and
$s'$, respectively.

We have $SO(3)$-invariance of the distances under the transformation
\[
(\Lambda_1(s),\br(s),\Lambda_2(s), \ldots, \Lambda_N(s))\mapsto (h\Lambda_1(s),h\br(s),h\Lambda_2(s)h^{-1}, \ldots , h\Lambda_N(s)h^{-1}) 
\]
for any $h \in SO(3)$, which shows that $SO(3)$ acts on the iterated semidirect product
\[
SE(3) \,\circledS\, (SO(3) (\,\circledS\, \cdots SO(3)) \cdots )
\quad\hbox{with $(N-1)$ copies of $SO(3)$.}
\] 
We apply 
the theory developed in \S\ref{sec:generalization} for $G_1 = \mathcal{F}(I,SE(3))$, $G_2 = \ldots = G_N = 
\mathcal{F}(I, SO(3))$. Thus the
Lagrangian variables are
\[
(\Lambda_1,\br,\Lambda_2,...,\Lambda_N)\in \mathcal{F}(I,SE(3)\,\circledS\,( SO(3)\,\circledS\,(...\,\circledS\, SO(3))))
\]
and the reduced variables are
\begin{equation}
\label{reduced_variables_list_N}
\begin{array}{llllll}
\omega_1&=&\Lambda_1^{-1}\dot\Lambda_1
&
\Omega_1&=&\Lambda_1^{-1}\Lambda'_1\\
\bgam&=&\Lambda_1^{-1}\dot\br
&
\bGam&=&\Lambda_1^{-1}\br'\\
\omega_2&=&\Lambda_1^{-1}\Lambda_2^{-1}\dot{\Lambda}_2\Lambda_1 \qquad \qquad\qquad \qquad
&
\Omega_2&=&\Lambda_1^{-1}\Lambda_2^{-1}\Lambda'_2\Lambda_1\\
\om_3&=&\Lambda_1^{-1}\Lambda_2^{-1}\Lambda_3^{-1}\dot{\Lambda}_3\Lambda_2\Lambda_1&
\Om_3&=&\Lambda_1^{-1}\Lambda_2^{-1}\Lambda_3^{-1}\Lambda_3'\Lambda_2\Lambda_1\\
...& ...&  &...& ...&\\
\om_N&=&\Lambda_1^{-1}\cdots\Lambda_N^{-1}\dot{\Lambda}_N\Lambda_{N-1}\cdots \Lambda_1 &
\Om_N&=&\Lambda_1^{-1}...\Lambda_N^{-1}\Lambda'_N\Lambda_{N-1}\cdots \Lambda_1\\
p_2&=&\Lambda_1^{-1}\Lambda_2\Lambda_1&
\brho&=&\Lambda_1^{-1}\br.\\
p_3&=&\Lambda_1^{-1}\Lambda_2^{-1}\Lambda_3\Lambda_2\Lambda_1\\
...&\\
p_N&=&\Lambda_1^{-1}...\Lambda_N...\Lambda_1
&
\end{array}
\end{equation}

\begin{remark} \rm 
\label{N-bouquet-notation} 
In the general case discussed above, there will be indices associated with $\Lambda_i^{k_{i-1}}$ labeling the attachment point of  each bouquet frame. Then, the reduced variables in \eqref{reduced_variables_list_N} will acquire 
indices as well, for example 
\begin{align*} 
&p_2^{k_1}=\Lambda_1^{-1}\Lambda_2^{k_1} \Lambda_1  \\
&p_3^{k_1 k_2} =\Lambda_1^{-1}\big( \Lambda^{k_1}_2\big) ^{-1} \Lambda_3^{k_2} \Lambda_2^{k_1} \Lambda_1\\
&...\\
&p_N^{k_1 k_2 \ldots k_{N-1}} =\Lambda_1^{-1} \ldots \Lambda_N^{k_{N-1}} \ldots \Lambda_1 \, . 
\end{align*} 
As we discussed earlier, this indexing poses no fundamental mathematical difficulties, but encumbers the notation in the resulting formulas. For the sake of simplicity and clarity of the exposition, we shall avoid this indexing in our further discussions. \quad $\blacklozenge$
\end{remark} 
The material Lagrangian depends only on
$(\Lambda_1,\dot{\Lambda}_1, \boldsymbol{r}, 
\dot{\boldsymbol{r}},\Lambda_2,\dot{\Lambda}_2, \ldots,
\Lambda_N,\dot{\Lambda}_N)$. As in the case of $2$-bouquets
the Lagrangian is the
restriction of a parameter-dependent Lagrangian defined
on $T \mathcal{F}(I, SE(3) \,\circledS\,(SO(3) \,\circledS\,(\cdots \,\circledS\,SO(3)\cdots)))\times 
\mathcal{F}(I, \mathbb{R}^3)^{N+2}$ for the zero value of the
parameter in $\mathcal{F}(I, \mathbb{R}^3)^{N+2}$, that is,
\[
L_{(a_0=0)} = L(\Lambda_1,\dot{\Lambda}_1, \boldsymbol{r}, 
\dot{\boldsymbol{r}},\Lambda_2,\dot{\Lambda}_2, \ldots,
\Lambda_N,\dot{\Lambda}_N).
\]
We are thus in the situation described by the theory developed in \S\ref{sec_fixed_par}, provided $L $
is $(G _1)^c_0$-invariant. We have $(G _1)^c_0= SO(3)
\subset \mathcal{F}(I, SE(3))$ as a direct verification
using \eqref{concrete_cocycle} shows, when extended to $N $ copies of $SO(3)$.

The most general form of the symmetry-reduced Lagrangian for $N $-bouquets is 
\begin{align*}\ell=&\ell_{loc}(\bom_1, \bgam, \bom_2,\ldots, \boldsymbol{\omega}_N, p_2,\ldots p_N,\bOm_1,\bGam,\bOm_2,\ldots, \boldsymbol{\Omega}_N, \brho)\\
& \quad + 
\ell_{np}(\bom_1, \bgam, \bom_2, \ldots, \boldsymbol{\omega}_N, p_2,\ldots p_N,\bOm_1,\bGam,\bOm_2,\ldots, \boldsymbol{\Omega}_N, \brho, (\Lambda_1, \br)),
\end{align*}
where $\ell_{loc}$ is explicitly given in terms of 
$(\bom_1, \bgam, \bom_2,\ldots, \boldsymbol{\omega}_N, p_2,\ldots p_N,\bOm_1,\bGam,\bOm_2,\ldots, \boldsymbol{\Omega}_N, \brho)$ and $\ell_{np}$ has still a dependence on $(\Lambda_1, \br)$, where $(\Lambda_1, \br)$ are such that \eqref{reduced_variables_list_N} holds. More 
precisely, $\ell_{np}$ has
the nonlocal expression
\begin{align*}
\ell_{np} = \iint U \big(\xi(s,s'), \bkappa(s,s'), p_2(s), \ldots, p_N(s), p_2(s'),\ldots, p_N(s'), \bGam(s), \bGam(s')\big) \mbox{d} s \,\mbox{d} s', 
\end{align*} 
where $U: SO(3) \times (\mathbb{R}^3)^{N+1} \rightarrow 
\mathbb{R}$ is a given function and $\bkappa(s,s') \in \mathbb{R}^3$, $\xi(s,s') \in SO(3)$ have the same formulas
as for the $2$-bouquets.

Given a Lagrangian $\ell$, formula \eqref{equations_N_bouquets} yields the following motion equations for $N$-bouquets comprising $(N+1)$ equations:
\begin{equation}
\left\{
\begin{array}{l}
\vspace{0.2cm}\displaystyle\left(\partial_t+\bom_1\times\right)\frac{\delta \ell}{\delta\bom_1}+\sum_{k=2}^N\bom_k\times \frac{\delta \ell}{\delta\bom_k}+\left(\partial_s+\bOm_1\times\right)\frac{\delta \ell}{\delta\bOm_1}+\sum_{k=2}^N\bOm_k\times \frac{\delta \ell}{\delta\bOm_k}\\
\vspace{0.2cm}\displaystyle \qquad\qquad +\brho\times\dede{l}{\brho}+\bGam\times\dede{l}{\bGam}+\bgam\times\dede{l}{\bgam}+
\sum_{k=2}^N  
\left(\frac{\delta\ell}{\delta p_k}p_k^{-1}-p_k^{-1}\frac{\delta\ell}{\delta p_k}\right)
=0\\
\vspace{0.2cm}\displaystyle\lp \prt_t + \bom_1\times\rp\dede{l}{\bgam} + \lp\prt_s + \bOm_1\times\rp\dede{l}{\bGam}
 -\dede{l}{\brho}=0\\
\qquad \vdots \qquad  \qquad \qquad \qquad \qquad \vdots\\
\vspace{0.2cm}\displaystyle \left(\partial_t+\bom_i\times\right)\frac{\delta \ell}{\delta\bom_i}+\sum_{k=1}^{i-1}\bom_k\times \frac{\delta \ell}{\delta\bom_i}+\sum_{k=i+1}^N\bom_k\times \frac{\delta \ell}{\delta\bom_k}\\
\qquad\qquad+\displaystyle\left(\partial_s+\bOm_i\times\right)\frac{\delta \ell}{\delta\bOm_i}+\sum_{k=1}^{i-1}\bOm_k\times \frac{\delta \ell}{\delta\bOm_i}+\sum_{k=i+1}^N\bOm_k\times \frac{\delta \ell}{\delta\bOm_k}\\
\qquad\qquad\qquad\qquad\displaystyle-
 p_i^{-1}\frac{\delta\ell}{\delta p_i}  
+\sum_{k=i+1}^N  \left(\frac{\delta\ell}{\delta p_k}p_k^{-1}-p_k^{-1}\frac{\delta\ell}{\delta p_k}\right)=0\\
\qquad \vdots \qquad  \qquad \qquad \qquad \qquad \vdots\\
\displaystyle\left(\partial_t+\bom_N\times\right)\frac{\delta \ell}{\delta\bom_N}+\sum_{k=1}^{N-1}\bom_k\times \frac{\delta \ell}{\delta\bom_N}\\
\qquad\qquad+\displaystyle\left(\partial_s+\bOm_N\times\right)\frac{\delta \ell}{\delta\bOm_N}+\sum_{k=1}^{N-1}\bOm_k\times \frac{\delta \ell}{\delta\bOm_N}-
p_N^{-1}\frac{\delta\ell}{\delta p_N}  =0
\end{array}
\right. 
\label{Nbouquet}
\end{equation}
where all the partial functional derivatives are total
derivatives. Note that $\frac{\delta \ell}{\delta p_i}(s) 
\in T\,^*_{p_i(s)} SO(3) $ and hence $p_i(s)^{-1}
\frac{\delta \ell}{\delta p_i}(s) \in \mathfrak{so}(3)^*
\cong \mathbb{R}^3$.
As in the case of $2$-bouquets these equations can be
made explicit and one gets an analogue of the system
\eqref{dynamic}.

\section{The equations of motion in conservative form} 
 
In order to elucidate the physical meaning of dynamic equations \eqref{compact_version_general}, we shall recast them in conservative form. In principle, the form of these equations express the laws of conservation of linear and angular momenta at the every branch level. However, the direct formulation in terms of these physical quantities leads to very cumbersome expressions, as the linear and angular motion of each branch contributes to the motion of all branches on the next levels, and the conservation laws for different bouquet levels are thus highly intertwined. It is preferable to use the geometric 
formulation and express the conservation laws as coadjoint motion on the corresponding semidirect product groups. In that case, the formulas become highly compact and yet their physical meaning remains clear. This geometric approach also lets us formulate the \emph{sequential conservation laws} in \S\ref{sec:sequential-conservation} below. 

\subsection{Two-bouquets}
 
Using \eqref{compact_version_general} and 
\eqref{adstar_sdp} , we see that the equations of motion 
\eqref{dynamic} take the form 
\begin{align}
\label{compact_equations}
&\left(\frac{\partial}{\partial t} - {\rm ad}^*_{(\bom_1,\bgam,\bom_2)} \right) 
\left(\frac{\de \ell}{\de \bom_1}, \frac{\de \ell}{\de \bgam},\frac{\de \ell}{\de \bom_2} \right)
+
\left(\frac{\partial}{\partial s} - {\rm ad}^*_{(\bOm_1,\bGam,\bOm_2)} \right)
\left(\frac{\de \ell}{\de \bOm_1}, \frac{\de \ell}{\de \bGam},\frac{\de \ell}{\de \bOm_2} \right) \nonumber \\
& \qquad \qquad  \qquad =
\left(\frac{\de \ell}{\de \brho} \times \brho, \frac{\de \ell}{\de \brho} , \mathbf{0} \right) 
+ \left(p^{-1} \frac{\delta \ell}{\delta p} 
- \frac{\delta \ell}{\delta p}p^{-1}, {\bf 0}, 
p^{-1}\frac{\delta \ell}{\delta p} \right).
\end{align} 
Therefore, we can write these equations in conservative form by using \eqref{conservation_general}. We get 
\begin{align} 
\label{conservation}
&\frac{\partial}{\partial t} {\rm Ad}^*_{(\Lambda_1, \br, \Lambda_2)^{-1}} 
\left(\frac{\de \ell}{\de \bom_1}, \frac{\de \ell}{\de \bgam},\frac{\de \ell}{\de \bom_2} \right)
+ 
\frac{\partial}{\partial s} {\rm Ad}^*_{(\Lambda_1, \br, \Lambda_2)^{-1}} 
\left(\frac{\de\ell}{\de \bOm_1}, \frac{\de \ell}{\de \bGam},\frac{\de \ell}{\de \bOm_2} \right) \nonumber\\
&\qquad \qquad 
= {\rm Ad}^*_{(\Lambda_1, \br, \Lambda_2)^{-1}} 
\left[\left( 
\frac{\de \ell}{\de \brho} \times \brho, \frac{\de \ell}{\de \brho} , \mathbf{0}  \right)
+ \left(p^{-1} \frac{\delta \ell}{\delta p} 
- \frac{\delta \ell}{\delta p}p^{-1}, {\bf 0}, 
p^{-1}\frac{\delta \ell}{\delta p} \right) \right]. 
\end{align} 
\subsection{$N$-bouquets} 

For $N$ bouquets, we obtain from \eqref{conservation_law} the following conservative form
\begin{align}
\label{conservation_N}
&\frac{\partial }{\partial t}\operatorname{Ad}^*_{(\Lambda_1,\br,\Lambda_2,...,\Lambda_N)^{-1}}\left(\frac{\delta\ell}{\delta\bom_1},\frac{\delta\ell}{\delta\bgam},...,\frac{\delta\ell}{\delta\bom_N}\right)
+\frac{\partial }{\partial s}\operatorname{Ad}^*_{(\Lambda_1,\br \Lambda_2, ...,\Lambda_N)^{-1}}
\left(\frac{\delta\ell}{\delta\bOm_1},\frac{\delta\ell}{\delta\bGam},...,\frac{\delta\ell}{\delta\bOm_N}\right)
\nonumber \\
&\qquad
=\operatorname{Ad}^*_{(\Lambda_1,\br,\Lambda_2,...,\Lambda_N)^{-1}}\left(\frac{\delta\ell}{\delta\brho}\times\brho,\frac{\delta\ell}{\delta\brho},0,...,0\right) \nonumber \\  
&\qquad \qquad - \operatorname{Ad}^*_{(\Lambda_1,\br,\Lambda_2,...,\Lambda_N)^{-1}} \left(
\begin{array}{c}\displaystyle
\sum_{k=2}^N 
\left(\frac{\delta \ell}{\delta p _k} p_k^{-1} - p_k^{-1}
\frac{ \delta \ell}{ \delta p _k} \right)\\ 
\displaystyle{\bf 0}\\ \displaystyle
- p_2 ^{-1}\frac{\delta \ell}{\delta p _2}+ \sum_{k=3}^N 
\left(\frac{\delta \ell}{\delta p _k} p_k^{-1} - p_k^{-1}
\frac{\delta \ell}{ \delta p _k} \right)\\ \displaystyle
\vdots\\ \displaystyle
- p_i ^{-1}\frac{\delta \ell}{\delta p _i} + \sum_{k=i+1}^N 
\left(\frac{\delta \ell}{\delta p _k} p_k^{-1} - p_k^{-1}
\frac{\delta \ell}{ \delta p _k} \right)\\ \displaystyle
\vdots\\ \displaystyle
- p_N^{-1}\frac{\delta \ell}{\delta p _N}
\end{array}
\right).
\end{align}

\subsection{Sequential conservative forms} 
\label{sec:sequential-conservation}
It is interesting to cast the equations  in conservative form not just for the strand as a whole, but for each particular bouquet and all the corresponding sequential bouquets. As an everyday analogy, consider a tree with many branches swinging  because some force shakes the trunk. The conservative form for the whole tree sums all the momenta for each branch and proves that all those momenta would sum to zero, were it not for the external force acting on the trunk. 

Let us now introduce an imaginary cut to a tree branch, and 
try to apply the corresponding conservation laws. 
All the forces (and torques) starting from the branch and propagating 
above to smaller branches must sum to zero in a conservation 
law. The external force will be now applied by the tree to 
the base of this branch. We can now continue to higher and 
higher branches and, as long as the tree structure is 
completely sequential, we should find an exact conservation law 
for all the branches at the level $n=1,2, \ldots$ and higher. 

Comparing the conservation laws for $N$-bouquets 
\eqref{conservation_N} with the equations of motion
\eqref{Nbouquet} for $K$-bouquets, we find a new conservation law  
for the first $K$ branches in the $N$-bouquets
\begin{align*}
&\frac{\partial }{\partial t}\operatorname{Ad}^*_{(\Lambda_1,\br,\Lambda_2,...,\Lambda_K)^{-1}}\left(\frac{\delta\ell}{\delta\bom_1},\frac{\delta\ell}{\delta\bgam},...,\frac{\delta\ell}{\delta\bom_K}\right)
+\frac{\partial }{\partial s}\operatorname{Ad}^*_{(\Lambda_1,\br \Lambda_2, ...,\Lambda_K)^{-1}}
\left(\frac{\delta\ell}{\delta\bOm_1},\frac{\delta\ell}{\delta\bGam},...,\frac{\delta\ell}{\delta\bOm_K}\right)
\nonumber \\
&\qquad
=\operatorname{Ad}^*_{(\Lambda_1,\br,\Lambda_2,...,\Lambda_K)^{-1}}\left(\frac{\delta\ell}{\delta\brho}\times\brho,\frac{\delta\ell}{\delta\brho},0,...,0\right) \nonumber \\ 
& \qquad \qquad
- \operatorname{Ad}^*_{(\Lambda_1,\br,\Lambda_2,...,\Lambda_K)^{-1}}\left(\ldots , 
\sum_{j=K+1}^N \boldsymbol{\omega}_j \times 
\frac{\delta\ell}{ \delta \boldsymbol{\omega}_j} +
\boldsymbol{\Omega}_j \times 
\frac{\delta\ell}{ \delta \boldsymbol{\Omega}_j}, \ldots 
\right) \\
& \qquad \qquad
- \operatorname{Ad}^*_{(\Lambda_1,\br,\Lambda_2,...,\Lambda_K)^{-1}} \left(\mathfrak{F}_{1}, 
{\bf 0}, \mathfrak{F}_{2}, \ldots, \mathfrak{F}_{K}\right),
\end{align*}
where 
\[
\mathfrak{F}_i = - p_i ^{-1}\frac{\delta \ell}{\delta p _i} + \sum_{k=i+1}^N 
\left(\frac{\delta \ell}{\delta p _k} p_k^{-1} - p_k^{-1}
\frac{\delta \ell}{ \delta p _k} \right), \quad i=1, \ldots,
K \leq N
\]
with the convention that $p_1$ is absent. 

More generally, the conservation laws for the branches
$J, \ldots, K$, $1 < J\leq K \leq N$, in the $N$-bouquet is
\begin{align*}
&\frac{\partial }{\partial t}\operatorname{Ad}^*_{(\Lambda_J,...,\Lambda_K)^{-1}}\left(\frac{\delta\ell}{\delta\bom_J},...,\frac{\delta\ell}{\delta\bom_K}\right)
+\frac{\partial }{\partial s}\operatorname{Ad}^*_{(\Lambda_J, ...,\Lambda_K)^{-1}}
\left(\frac{\delta\ell}{\delta\bOm_J},...,\frac{\delta\ell}{\delta\bOm_K}\right)
\nonumber \\
& = 
- \operatorname{Ad}^*_{(\Lambda_J,...,\Lambda_K)^{-1}}
\left(\ldots , 
\sum_{j=K+1}^N \boldsymbol{\omega}_j \times 
\frac{\delta\ell}{ \delta \boldsymbol{\omega}_j} +
\boldsymbol{\Omega}_j \times 
\frac{\delta\ell}{ \delta \boldsymbol{\Omega}_j}, \ldots 
\right) \\
&\qquad \qquad
- \operatorname{Ad}^*_{(\Lambda_J,...,\Lambda_K)^{-1}}
\left(\ldots, 
\sum_{j=1}^{J-1} \boldsymbol{\omega}_j \times 
\frac{\delta\ell}{ \delta \boldsymbol{\omega}_i} +
\boldsymbol{\Omega}_j \times 
\frac{\delta\ell}{ \delta \boldsymbol{\Omega}_i}, \ldots 
\right)\\
& \qquad \qquad
- \operatorname{Ad}^*_{(\Lambda_J,...,\Lambda_K)^{-1}} 
\left(\mathfrak{F}_{J}, \ldots, \mathfrak{F}_{K}\right),
\end{align*}
For example, taking $J=K=N=2$ for $2$-bouquets yields
\begin{equation*} 
\frac{\partial }{\partial t} {\rm Ad}^*_{\Lambda_2^{-1}} 
\frac{\de l}{\de \bom_2} + 
\frac{\partial }{\partial s} {\rm Ad}^*_{\Lambda_2^{-1}} \frac{\de l}{\de \bOm_2}
= \underbrace{
-{\rm Ad}^*_{\Lambda_2^{-1}}\Big( 
\bom_1 \times \frac{\de l}{\de \bom_2}+
\bOm_1 \times \frac{\de l}{\de \bOm_2}
\Big)
}_{\mbox{1st bouquet acting on 2nd bouquet}}
+ \operatorname{Ad}_{ \Lambda_2^{-1}}^\ast \left(p ^{-1} 
\frac{\delta \ell}{ \delta p}\right).
\label{sequential} 
\end{equation*}
\section{Conclusion} 
As with many other complex biological and organic molecules, the final conformation of the compound polymers considered here cannot be determined only on the basis of the forces acting on the polymer. The choice of the dynamical path in configuration space taken by the molecule during conformation is also of crucial  importance.  
In this paper, we have outlined a geometrically exact theory of dendritic polymer dynamics based on symmetry reduction of Hamilton's principle defined on the tangent space of the iterated semidirect-product Lie group
\[
SE(3)\,\circledS\,(SO(3)\,\circledS\,(SO(3)\,\circledS\,(\cdots\,\circledS\, SO(3))\cdots))
\]
for Lagrangians that are invariant under changes of orientation $SO(3)$ in the first Lie group, $SE(3)\simeq SO(3)\,\circledS \,\mathbb{R}^3$. As in the example of shaking the trunk of a tree with many branches, the overall orientation and position of the tree may be  immaterial, but the relative orientations of its successive branches are important to the time-dependent  oscillations and deformations of the tree away from its reference configuration.  The motivation for the iterated semidirect product action was illustrated by considering the case of coupled rigid bodies, and was developed into a general theory of dendritic polymers with many branches and types of rigid charge bouquets. However, specific configurations and individual molecular parameters, such as moments of inertia and length-scales, were not considered. Although the approach was general, the calculations were performed explicitly enough for real applications of evolutionary modeling of dendritic polymers to
  be made, once their reference configurations and molecular parameters were specified.  We concentrated mainly on two-level bouquets, because this case illustrates most of the mathematical concepts and is the necessary first step in applications. However, formulas for $N$-level dendrimers were also derived and were discussed in enough detail to make their general pattern evident for understanding the scope of further applications. Both local and nonlocal interactions were included in the theory, in order to account for the screened electrostatic interactions that occur among different branches of the dendronized polymers. Finally, the theory was expressed in conservative form, so that forces and torques among the branches could be explicitly identified. 

While our theory is very general and is applicable to a wide range of elastic and nonlocal forces, the major drawback at this stage is that it does not consider dissipation due to possible presence of ambient fluid. Indeed, the most natural 
media for polymers is a liquid solution, where the forces of friction dominate the dynamics. Our theory does not consider the forces of friction and dissipation and thus is not readily applicable to polymers in liquids. It is not clear at this stage how to derive geometrically exact analytical expressions for friction forces acting on the dendrimers in an arbitrary conformation. 
Presumably, further progress in that field will need to include some insightful, yet ad-hoc, physical approximations for frictional forces as a function of local velocities in the semidirect product groups considered here. We did not dare to venture into this territory here and only considered the case when the friction effect due to the ambient media can be neglected.

\section{Acknowledgements} 
We thank M. Bruveris, D.C.P. Ellis, and J.E. Marsden for helpful discussions of this work as it was being developed. 
The work of FGB was partially supported by a Swiss NSF postdoctoral fellowship.
DDH and VP were partially supported by NSF grants NSF-DMS-05377891 and NSF-DMS-09087551. The work of
DDH was also partially supported  by the Royal Society of London Wolfson Research Merit Award.
TR was partially supported by a Swiss NSF grant and the program \textit{Symplectic and Contact Geometry and Topology} at MSRI Berkeley. 

\bibliographystyle{unsrt}
\bibliography{papers} 

\end{document}